\documentclass[]{aastex631}

\usepackage{amsmath}
\usepackage{longtable}

\newcommand{\htwo}{H$_2$}
\newcommand{\htwoo}{H$_2$O}
\newcommand{\otwo}{O$_2$}

\newcommand{\eminus}{e$^-$}

\accepted{\today}

\submitjournal{Icarus}

\shorttitle{NLTE modelling of {\htwoo}}
\shortauthors{Garc\'ia Mu\~noz et al.}

\begin{document}

\title{NLTE modelling of water-rich exoplanet atmospheres. 
Cooling and heating rates.}

\correspondingauthor{Antonio Garc\'ia Mu\~noz}
\email{antonio.garciamunoz@cea.fr, tonhingm@gmail.com}

\author[0000-0003-1756-4825]{A. Garc\'ia Mu\~noz}
\affiliation{
Universit\'e Paris-Saclay, Universit\'e Paris Cit\'e, CEA, CNRS, AIM, 91191, Gif-sur-Yvette, France}
\author[0000-0002-1248-0553]{A. Asensio Ramos}
\affiliation{
Instituto de Astrof\'isica de Canarias, 38205, La Laguna, Tenerife, Spain; 
Departamento de Astrof\'isica, Universidad de La Laguna, 38205, La Laguna, Tenerife, Spain}
\author[0000-0001-7199-2535]{A. Faure}
\affiliation{
Universit\'e Grenoble Alpes, CNRS, IPAG, F-38000 Grenoble, France}

\begin{abstract}
The hydrogen and water molecules respond very differently to the collisional-radiative 
processes taking place in planetary atmospheres.
Naturally, 
the question arises whether {\htwoo}-rich atmospheres are more (or less) resilient to 
long-term mass loss than {\htwo}-dominated ones 
if they radiate away the incident stellar energy more (or less) efficiently. 
If confirmed, the finding  
would have implications on our understanding of the evolution of sub-Neptune exoplanets.
As a key step towards answering this question, 
we present a non-local thermodynamic equilibrium (NLTE) model of
{\htwoo} for the atmospheric region where
the gas accelerates to escape the planet and conditions relevant to close-in sub-Neptunes.
Our exploratory calculations for isothermal gas composed of {\htwo}, {\htwoo} and {\eminus} reveal that:
$\textit{1}$) In the pressure region $\sim$10$^{-2}$--10$^{-4}$ dyn cm$^{-2}$ 
where the stellar extreme-ultraviolet (XUV) photons are typically deposited in the atmosphere, 
{\htwoo} is in rotational LTE but vibrational NLTE. 
Vibrational LTE is facilitated by high {\htwoo} abundances and fractional ionizations, 
and we report critical densities for the LTE-NLTE transition; 
$\textit{2}$) Vibrational cooling 
may locally dominate over rotational cooling, 
partly because of the comparatively small opacities of ro-vibrational lines;  
$\textit{3}$) Even low {\htwoo} abundances notably enhance the cooling, foreseeably 
offsetting some of the stellar heating; 
$\textit{4}$) Heating due to the deposition of stellar infrared (IR)
photons is significant at pressures $\gtrsim$0.1 dyn cm$^{-2}$. 
We estimate the contribution of {\htwoo} excitation to the internal energy of the gas
and speculate on the photodissociation from the excited vibrational states. 
Ultimately, our findings motivate the consideration of 
 NLTE in the mass loss rate calculations of {\htwoo}-rich atmospheres.

\end{abstract}


\vspace{4.5cm}

\section{\label{intro_sec} Introduction}

Water is thought to be abundant at some low-mass exoplanets 
\citep{venturinietal2020,bitschetal2021,kiteschaefer2021,kimuraikoma2022,izidoropiani2023}.
Proving the reality of {\htwoo}-rich planets is challenging though as 
mass and size alone, two properties 
accurately known for many planets, are often insufficient to conclude on their 
interior structures and atmospheric compositions \citep{valenciaetal2007,acunaetal2021,delrezetal2021}. 
Even when the transmission spectrum shows evidence for {\htwoo}, 
the intrinsic degeneracies in the interpretation of the spectrum make it difficult 
to decide whether {\htwoo} is a major or minor atmospheric constituent
\citep{bennekeetal2019,madhusudhanetal2020}. 
The long-term stability of the atmosphere provides a complementary way of
assessing its composition. 
There is indeed a growing sample of planets for which the existence of atmospheres is
firmly established but they are unlikely to be {\htwo}-dominated as they would
have been lost to space long ago under their current irradiation conditions. 
That leaves {\htwoo} and other gases
of high mean molecular weight as more plausible alternatives
\citep{garciamunozetal2020,garciamunozetal2021,piauletetal2023}.
\\

The community is in the process of testing these and other ideas about the 
atmospheric composition of low-mass planets with the James Webb Space Telescope (JWST)
\citep{ihetal2023,lincowskietal2023,moranetal2023} and other facilities. 
Simultaneously, we must assess the validity of
the assumptions made in the models used for predicting the mass loss rates and how they 
affect the stability of the atmospheres. 
For example, a {\htwoo}-rich atmosphere is expected to respond to infrared (IR)
radiation very differently than a {\htwo}-dominated one because, unlike the {\htwo} molecule, 
{\htwoo} has a large dipole moment. 
Also, {\htwoo} has numerous low-energy states 
that are easily populated in collisions with molecules (including atoms)
and electrons and that, once excited, radiate promptly. 
In particular, self-collisions with other {\htwoo} molecules are efficient at
populating the {\htwoo} states.
\\

Our work assesses the significance of bound-bound collisional-radiative processes within the {\htwoo} molecule 
for the cooling and heating (net cooling, hereafter) of {\htwoo}-rich atmospheres.
We build for that purpose a model that solves the Non-Local Thermodynamic Equilibrium (NLTE) 
problem of {\htwoo} under quite general conditions. 
The NLTE treatment is warranted at the low pressures where the atmospheric gas
is accelerated outwards to escape the planet and the population of {\htwoo} states 
may more easily depart from a Boltzmann distribution. 
The model considers the collisions of {\htwoo} with the background gas, and the 
interaction with upwelling radiation emitted from deeper atmospheric layers, 
with internally-generated diffuse radiation and with stellar radiation. 
The radiative interactions occur in the far-IR
for purely rotational transitions (that occur with no change in the vibrational quantum numbers) 
and in the near- and mid-IR for ro-vibrational transitions (that occur with changes in the vibrational quantum numbers). 
To our knowledge, a similar investigation of exoplanet atmospheres has never been done.  
We are exploring the implications on various atmospheric properties
and on the demographics of sub-Neptunes in ongoing work. 
The connected nature of the energy budget, chemistry and dynamics 
of atmospheres \citep{sanchez-lavegaetal2023} suggests that {\htwoo} NLTE might alter the three-dimensional structure
of exoplanet atmospheres, especially at low and intermediate pressures. 
The idea remains to be investigated by existent General Circulation Models
\citep{kaspishowman2015,wordsworth2015,caroneetal2018}.
\\

The {\htwoo} NLTE problem is certainly not new in the planetary 
sciences and astrophysics. 
It has been treated in the remote sensing of the
Earth's atmosphere to constrain its temperature and energy budget 
and to infer {\htwoo} abundances  \citep{lopez-puertastaylor2001,feofilovetal2009}.
It has been considered in interstellar and 
circumstellar clouds, stars and galaxies  \citep{cernicharocrovisier2005,vandishoecketal2013,grayetal2016}, 
prompting various calculations of cooling rates
\citep{goldsmithlanger1978,hollenbachmckee1979,neufeldkaufman1993,morrisetal2009}. 
In the latter astrophysical environments, {\htwoo} occurs in trace amounts and its collisional excitation 
is dominated by {\htwo} and, possibly, H atoms or electrons. 
The {\htwoo} NLTE problem has also been treated in the investigation
of the envelopes of comets and icy moons \citep{crovisier1984,xiemumma1992,
gerschetal2018,villanuevaetal2023}, where {\htwoo} is a major constituent.
These works often focus on predicting the 
populations of excited states for comparison against remote sensing measurements. 
The problem addressed here is somewhat different and more general. 
We consider a broad range of compositions, from {\htwoo} being a trace constituent 
to being dominant, and the possibility of both the
purely rotational and ro-vibrational excitation of {\htwoo} in collisions with {\htwo}, {\htwoo} and electrons
of kinetic temperatures from 200 to 1,500 K. 
Our work focuses on the cooling rates rather than on the specifics of the {\htwoo} spectrum.

\section{The {\htwoo} molecule}

{\htwoo} occurs in ortho (o--{\htwoo}; total nuclear spin of the 
hydrogen atoms $I$=1; nuclear spin statistical weight $g_I$=3) and para  (p-{\htwoo}; $I$=0; $g_I$=1) isomers.
The two isomers do not readily interconvert 
through collisions or radiation, and we treat them separately.
As a non-linear triatomic molecule, {\htwoo} possesses 3
 vibrational modes of motion, identified by the  quantum numbers
$v$=($v_1 v_2 v_3$)
for symmetric stretching ($v_1$), bending ($v_2$), and asymmetric stretching ($v_3$). 
The rotational motion is described by the quantum numbers ($J K_a K_c$), 
or alternatively ($J_{\tau}$ with $\tau${$\equiv$}{$K_a${$-$}$K_c$}). 
$J$ is the total rotational quantum number, and $K_a$ and $K_c$ are the respective projections 
of $J$ on the molecular axes with the smallest and greatest moments of inertia.
\\

Building the NLTE model requires specifying the participating o- and p-{\htwoo} states, 
the collisional-radiative processes through which they interact with the medium  
and their rate coefficients. 
Full details on our molecular model 
are given in Appendix \ref{molecularmodel_appendix}. 
In short, it consists for o-{\htwoo} of {$\sim$}400 states within the five lowest-energy
vibrational states, namely  $v$=(000), (010), (020), (100) and (001).
They are connected through {$\sim$}7,600 radiative transitions. 
It considers the collisions of o-{\htwoo} with the molecules {\htwo} and {\htwoo}
and with electrons. 
About 83,000 channels for 
inelastic collisions of o-{\htwoo} with electrons are included, 
and comparable numbers for collisions with {\htwo} and {\htwoo}.
Inelastic collisions with ions are omitted but this should not be a limitation of the
model as the ion densities in the atmosphere are small where the {\htwoo} molecule 
remains undissociated, 
and inelastic collisions with electrons are dominant once ionization becomes significant.
The number of states, radiative transitions and collisional channels for 
p-{\htwoo} are similar to those for o-{\htwoo}.
\\

\section{Cooling rate}

We review the formulation of the NLTE problem and its contribution to the net cooling 
of the gas. The treatment is standard and we include it here for later reference.\\

We are concerned with the energy equation for a radiating fluid \citep{zeldovichraizer2002}:
\begin{equation}
\frac{\partial \rho E}{\partial t} + \nabla \cdot \left[ \rho  H \mathbf{u} + \mathbf{q}  \right]
- \rho \mathbf{f}_{\rm{ext}} \cdot \mathbf{u}
=-\Gamma=-\int \frac{dI_{\lambda}}{ds} d\Omega d\lambda.
\label{energy_equation}
\end{equation}
Here, $\rho E$ is the total energy, including the bulk kinetic 
energy and the internal energy; $\rho H$=$\rho E+p$ is the total enthalpy, and 
$p$ is the pressure; $\mathbf{u}$ is the bulk velocity of the gas, and  
$\rho \mathbf{f}_{\rm{ext}} \cdot \mathbf{u}$ is the work of the external forces;
$\mathbf{q}$ is the heat flux. 
The internal energy contains the enthalpy of formation (plus dissociation and ionization)
of the different particles (molecules and electrons), 
and their translational, rotational, vibrational and electronic energies (as applicable).
\\

The right hand side of Eq. \ref{energy_equation} represents the exchange of energy  
between the gas and its surroundings through radiation. 
$\Gamma$ is calculated by solving the radiative transfer equation:
\begin{equation}
\frac{dI_{\lambda}}{ds}=-\kappa_{\lambda} I_{\lambda} + \varepsilon_{\lambda},
\label{rte_eq}
\end{equation}
which expresses the variation in radiance $I_{\lambda}$ per unit length $ds$, for 
absorption and emission coefficients $\kappa_{\lambda}$ and $\varepsilon_{\lambda}$,  
and integrating it over solid angle $\Omega$ and wavelength $\lambda$.
Radiation occurs over bound-bound, bound-free/free-bound and free-free transitions, 
and $\kappa_{\lambda}$=$\kappa_{\lambda}^{\rm{BB}}$+$\kappa_{\lambda}^{\rm{BF}}$+$\kappa_{\lambda}^{\rm{FF}}$ 
and $\varepsilon_{\lambda}$=$\varepsilon_{\lambda}^{\rm{BB}}$+$\varepsilon_{\lambda}^{\rm{FB}}$+$\varepsilon_{\lambda}^{\rm{FF}}$, 
such that:  
\begin{equation}
\frac{dI_{\lambda}}{ds}=\underbrace{-\kappa^{\rm{BB}}_{\lambda} I_{\lambda} + \varepsilon^{\rm{BB}}_{\lambda}}_{\rightarrow \Gamma^{\rm{BB}}} 
\underbrace{-\kappa^{\rm{BF}}_{\lambda} I_{\lambda} + \varepsilon^{\rm{FB}}_{\lambda}}_{\rightarrow \Gamma^{\rm{BF/FB}}} 
\underbrace{-\kappa^{\rm{FF}}_{\lambda} I_{\lambda} + \varepsilon^{\rm{FF}}_{\lambda}}_{\rightarrow \Gamma^{\rm{FF}}}, 
\end{equation}
and separate contributions $\Gamma^{\rm{BB}}$, $\Gamma^{\rm{BF/FB}}$ and $\Gamma^{\rm{FF}}$ can be defined.
In practice, 
the solution to the radiative transfer problem for the bound-bound transitions can be decoupled 
from the general radiative transfer problem if at the relevant wavelengths  
$\kappa_{\lambda}${$\approx$}$\kappa^{\rm{BB}}_{\lambda}$ and $\varepsilon_{\lambda}${$\approx$}$\varepsilon^{\rm{BB}}_{\lambda}$.
This is the case for the problem of interest here because the bound-bound radiative 
transitions of {\htwoo} occur mostly in the IR, 
whereas the bound-free/free-bound transitions occur in the ultraviolet and the free-free transitions typically
contribute weakly across the broad spectrum. 
$\Gamma^{\rm{BF/FB}}$ and $\Gamma^{\rm{FF}}$ are often incorporated in the
hydrodynamical models that solve Eq. \ref{energy_equation}, whereas 
$\Gamma^{\rm{BB}}$ is omitted or severely simplified in them.
Our focus in this work is to calculate $\Gamma^{\rm{BB}}$ 
for a variety of conditions relevant to {\htwoo}-rich atmospheres.
\\

The absorption and emission coefficients for bound-bound transitions are given by sums 
over all the {\htwoo} states of indices $i$, $j$,  
(number) densities $n_i$, $n_j$ and energies $E_j${$>$}$E_i$ that are connected radiatively: 
\begin{equation}
\kappa^{\rm{BB}}_{\lambda}=\sum_{i,j>i} \frac{h \lambda_{ij}}{4\pi} ( n_i B_{ij}  - n_j B_{ji}  ) \psi^{ij}_{\lambda} 
\label{kappabb_eq}
\end{equation}
\begin{equation}
\varepsilon^{\rm{BB}}_{\lambda}=\sum_{i,j>i} \frac{hc}{4\pi\lambda_{ij}} 
n_j A_{ji} \psi^{ij}_{\lambda} %\;\Big[\frac{\text{erg}}{\text{cm}^3 \text{ s sr cm}}\Big]
\end{equation}
where $h$ and $c$ are the Planck constant and speed of light, respectively, and 
$\lambda_{ij}$=$hc$/($E_j${$-$}$E_i$) is the wavelength. 
$A_{ji}$, $B_{ji}$ and $B_{ij}$ are the Einstein coefficients for spontaneous emission, induced
emission and absorption, related through $B_{ji}$/$A_{ji}$=$\lambda^3_{ij}$/$2hc$ and
$g_j B_{ji}$=$g_i B_{ij}$. 
The above expressions assume complete frequency redistribution, with absorption and emission 
described by the same line profile $\psi^{ij}_{\lambda}$.\\

Following the above, 
the net cooling rate for bound-bound transitions, or net cooling
rate for short, is given by:
\begin{equation}
\Gamma^{\rm{BB}}=
\sum_{i,j>i} \left[ n_j A_{ji} - ( n_i B_{ij}-n_j B_{ji} ) \Phi_{ij} \frac{\lambda^2_{ij}}{c} \right] (E_j - E_i),
\label{gammabb_eq}
\end{equation}
where:
\begin{equation*} 
\Phi_{ij} = \frac{1}{4 \pi}  \int \int I_{\lambda} \psi^{ij}_{\lambda} d\Omega d\lambda 
\end{equation*}
is the mean line intensity integrated over wavelength. 
Alternatively:
\begin{equation}
\Gamma^{\rm{BB}}=
\sum_{i,j>i} n_j A_{ji} p_{ji}  (E_j - E_i),
\label{gammabb_radbracket_eq}
\end{equation}
and the so-called net radiative bracket \citep{athayskumanich1971,elitzurasensioramos2006}:
\begin{equation}
p_{ji}= 1 - \frac{n_i g_j - n_j g_i}{n_j g_i} \frac{\lambda^5_{ij} \Phi_{ij}}{2hc^2}.
\end{equation}
As used here, $p_{ji}$ includes in $\Phi_{ij}$ the contributions from
 both external radiation entering the atmosphere
and diffuse radiation generated within. 
We note however that the usual definition of $p_{ji}$ includes only the contribution
from diffuse radiation \citep{elitzurasensioramos2006}, 
in which case it is related to the probability for the line photon to escape the gas
\citep{athayskumanich1971,irons1978}, 
whereas the external contribution is treated through a separate term. 
This being a matter of nomenclature that affects only the calculations for which the gas 
is externally irradiated, 
we have followed the above definition because the corresponding $p_{ji}$ are easy to extract from our
NLTE solver and because keeping separate terms does not add any physical insight to
the discussion.
It is sometimes useful to separate $\Gamma^{\rm{BB}}$ into the 
components $\Gamma^{\rm{BB}}_{v}$, each of them collecting the cooling rate from 
radiative transitions with upper states in the specified vibrational state $v$.
By construction, the summation of the 
rotational component $\Gamma^{\rm{BB}}_{000}$ and the ro-vibrational components 
$\Gamma^{\rm{BB}}_{010}$, $\Gamma^{\rm{BB}}_{020}$, $\Gamma^{\rm{BB}}_{100}$ and 
$\Gamma^{\rm{BB}}_{001}$ amounts to the net cooling rate. 
\\

Figure \ref{H2O_sketch_fig} sketches the {\htwoo} molecular model.  
The system is assumed closed, meaning that any collisional-radiative process that 
initiates from a bound state $i$ results into another bound state $j$. 
This generally requires that the chemical processes producing or destroying 
the {\htwoo} molecule proceed more slowly than the relaxation within it,
thereby rendering the two problems separable. 
This condition is approximately met in the region of an
atmosphere where {\htwoo} remains undissociated and where the relaxation 
times are shorter than the chemical loss times.
For our NLTE calculations, 
we have assumed that the total density of {\htwoo} is known and satisfies 
[{\htwoo}]={$\sum_i$}{$n_i$}.\\

The density of a state $i$ is determined by solving the steady-state problem: 
\begin{equation}
\frac{dn_i}{dt}=0=\sum_{j > i}  n_j A_{ji} p_{ji}
-
\sum_{j < i}  n_i A_{ij} p_{ij}
+ \sum_{j \neq i} (n_j C_{ji} - n_i C_{ij}).
\label{dnidt_eq}
\end{equation}
for the balance of production and loss rates through all possible collisional-radiative processes. 
$C_{ji}$ is the frequency for collisional deexcitation from $j$ to $i$ for
$j${$>$}{$i$}, and the frequency for collisional excitation from $j$ to $i$ for 
$j${$<$}{$i$}. 
Summation of this equation $\times${$E_i$} over all the states results in the equivalent
form for the net cooling rate:
\begin{equation}
\Gamma^{\rm{BB}}= \sum_{i,j>i}(n_i C_{ij}-n_j C_{ji}) (E_j - E_i).
\label{gammabb_coll_eq}
\end{equation}
Cooling ($\Gamma^{\rm{BB}}${$>$}0) occurs when excitation dominates over deexcitation, and
heating ($\Gamma^{\rm{BB}}${$<$}0) when the reverse is true.

\section{Numerical solver: MOLPOP-CEP}

In NLTE, the densities of the {\htwoo} states depart from 
a Boltzmann distribution at the kinetic temperature of the gas. 
Solving the NLTE problem requires solving the equivalent of Eq. \ref{dnidt_eq} for the 
densities jointly with Eq. \ref{rte_eq} for the radiation field.
We have used for that the publicly-available MOLPOP-CEP code \citep{asensioramoselitzur2018}, 
built upon the Coupled Escape Probability (CEP) formalism that tackles exactly the joint
population-radiation problems \citep{elitzurasensioramos2006}. 
MOLPOP-CEP outperforms the $\Lambda$-iteration 
technique for short characteristics and parabolic interpolation of the source function, 
which is a standard NLTE solver for benchmarking. 
For example, Tables 1-2 and Figures 3-6 in \citet{elitzurasensioramos2006} quantify 
both the computational cost and error incurred by both approaches for a variety of
configurations, showing that for a comparable computational cost the 
error associated with the MOLPOP-CEP calculations is typically orders of magnitude 
lower. 
Similar findings have been reported by \citet{yunetal2009} in a comparison of the CEP 
formalism generalized to spherical-shell atmospheres against Monte Carlo simulations.
\\

MOLPOP-CEP requires as input the total density of the target molecule, here
o- or p-{\htwoo}.
It also requires the list and densities of the colliders for collisions 
with the target molecule ({\htwo}, {\htwoo} and $e^-$ in our application). 
The model implicitly assumes that the velocities and state-resolved densities of the 
colliders do not significantly depart from Maxwell-Boltzmann distributions and that the 
collisional properties for each target-collider pair can be described by a kinetic 
temperature $T$. We will relax some of these simplifications in future work to include for example the 
important effect of non-thermal electrons in the net cooling of the atmosphere 
\citep{garciamunoz2023a,garciamunoz2023b,gilletetal2023}.
The above properties are specified over a spatial grid of coordinate $z$
that represents the altitude in the atmosphere. 
As boundary conditions, MOLPOP-CEP requires the external radiation entering
the simulation domain.  
We have considered external irradiation from above and below to mimic 
the radiation from the star hosting the planet and the radiation upwelling from deeper layers of the atmosphere. 
We have modified the public version of MOLPOP-CEP by extending the collisional 
and radiative properties in the model.
We have also modified the code to write out the terms that appear in Eqs. 
\ref{gammabb_radbracket_eq} and \ref{gammabb_coll_eq}, with which we calculate the
net cooling rate.
The Appendix and Supplementary Information give extended accounts of how MOLPOP-CEP was
adapted to our application.
\\

For completeness, we note some potential caveats in the solution of the {\htwoo} NLTE 
problem with MOLPOP-CEP.
Firstly, MOLPOP-CEP assumes a plane-parallel atmosphere. The simplification is likely not critical
as we are mainly concerned with a narrow atmospheric layer where the {\htwoo} molecule remains undissociated.
Secondly, 
MOLPOP-CEP describes the radiative transitions of the two {\htwoo} isomers
by non-overlapping Gaussian line shapes. 
The simplification is reasonable at low pressures for which the line wings remain narrow. 
Thirdly, MOLPOP-CEP cannot accommodate near-resonant 
channels such as {\htwoo}(010)+{\otwo}(0)$\leftrightarrow${\htwoo}(000)+{\otwo}(1)
that are important in {\otwo}-rich atmospheres \citep{yankovskyetal2011,funkeetal2012,langetal2020}. 
We assume that such collisional channels are not dominant 
in the atmospheres that we are exploring, an idea that can always be assessed \textit{a posteriori}. 
Lastly, MOLPOP-CEP assumes a static atmosphere. 
In principle, this simplification may prevent the escape of radiation that occurs
when the emitting and absorbing lines become mutually Doppler-shifted in an
accelerating gas. 
We argue  in {\S}\ref{static_section} though that this simplification is unlikely to cause significant errors 
in the net cooling of the region where the {\htwoo} molecule remains undissociated.

\section{Some estimates}

Equation \ref{gammabb_radbracket_eq} is the basis for our discussion
on the net cooling rates. 
Some insight into it can be gained with a simple two-state molecular model, 
for which:
\begin{eqnarray*}
n_1 + n_2 & = & [{\rm{H_2O}}]   \\
\frac{n_2}{n_1} & = & \frac{C_{21}}{A_{21}p_{21}+C_{21}} \frac{g_2}{g_1} e^{-(E_2-E_1)/kT} \\
\Gamma^{\rm{BB}} & = & n_2 A_{21} p_{21} (E_2 - E_1), 
\end{eqnarray*}
where we have used detailed balancing to relate the collisional frequencies.
Frequent collisions drive the $n_2$/$n_1$ ratio towards a Boltzmann distribution, 
with $C^{*}_{21}${$\sim$}$A_{21}p_{21}$ setting the condition on the critical density 
of the background gas and $C_{21}${$\gg$}$C^{*}_{21}$ the condition at which LTE is attained. 
The net radiative bracket
$p_{21}$ contains information on two non-local phenomena, namely 
 line opacity and the irradiation of the gas by photons emitted elsewhere. 
Assuming  the first of these dominant, $p_{21}$ is often interpretable
as the probability that the line photons may escape from their local
environment \citep{irons1978}, in which case $p_{21}${$\ll$1} if the line is optically
thick and  $\sim$1 if optically thin. 
In these limits, it is said that self-absorption is strong and weak, respectively.
A small $p_{21}$ helps drive the states towards LTE, possibly ensuring a large 
$n_2$/$n_1$ ratio. 
As a competing effect, 
a small $p_{21}$ causes the attenuation of the net cooling rate by the effective 
reduction of the transition probability. 
These effects combine very differently in purely rotational and in ro-vibrational lines.
\\

Purely rotational transitions between two states with $v$=(000) 
are typically characterized by moderate transition probabilities, small energy differences
and large densities. For them 
$p_{21}$ becomes $\ll$1 in the moderately dense layers of the atmosphere. 
Ro-vibrational transitions have in comparison 
large transition probabilities and energy differences, but the density $n_2$
is likely small if the state is populated through collisions and 
the temperatures remain moderate, i.e. if ({$E_2$}{$-$}$E_1$){/$kT$}{$\gg$1}. 
The line opacities associated with ro-vibrational transitions are generally much smaller
than those of purely rotational transitions and their net radiative brackets are closer to 1.

\subsection{\label{ndcrit_subsec}Critical densities for rotational and vibrational thermalization}

We use the data compiled in Appendix \ref{molecularmodel_appendix}
(the band averages of the transition probabilities and rate coefficients for deexcitation)
to estimate some critical densities for the background gas. 
We do it for an arbitrary $T$=400 K to offer some initial insight. 
Considering first rotational thermalization within $v$=(000), 
with $A_{v'v''}${$\sim$}1.5 s$^{-1}$, 
$k^{e^-}_{v'v''}${$\sim$}5$\times$10$^{-7}$ cm$^3$s$^{-1}$, 
$k^{\rm{H}_2}_{v'v''}${$\sim$}2.5$\times$10$^{-10}$ cm$^3$s$^{-1}$, 
$k^{\rm{H}_2\rm{O}}_{v'v''}${$\sim$}1.3$\times$10$^{-9}$ cm$^3$s$^{-1}$
and an \textit{ad-hoc} radiative bracket of 1, 
the critical densities are as low as 6$\times$10$^9$ and 1$\times$10$^9$ cm$^{-3}$ 
for neutral atmospheres of {\htwo} and {\htwoo}, respectively. 
For collisions with electrons, the critical density 
is 3$\times$10$^6$ cm$^{-3}$, well below the peak densities 
predicted in the upper atmospheres of many exoplanets \citep[see Fig. 2 of][]{garciamunozetal2021}. 
In such cases, 
the collisions with electrons ensure the rotational thermalization of
{\htwoo} to lower pressures than enabled by the collisions with neutrals alone.
\\

Listed in Table \ref{ndcrit_table} are 
the critical densities for the vibrational states $v$=(010), (020), (100) and (001).
We have estimated them by means of the generalized condition 
$\sum_{v'' < v'}A_{v'v''}${$\sim$}$\sum_{v''< v'}C_{v'v''}$.
Focusing on $v$=(010), the critical densities of {\htwo} and {\htwoo} 
are 1.5$\times$10$^{13}$ and 4.3$\times$10$^{11}$ cm$^{-3}$, respectively. 
Therefore
vibrational thermalization of the bending mode occurs at pressures lower by 1-2 orders of magnitude when the 
background gas is {\htwoo} than when it is {\htwo}. 
The corresponding critical density of electrons 
is 1.7$\times$10$^9$ cm$^{-3}$.
This is larger than the electron densities usually predicted for the upper atmospheres
of exoplanets. The contribution of the electrons to the vibrational thermalization of
{\htwoo} in the lower atmosphere may also be minor, as the overall fractional 
ionization decreases towards the lower atmospheric layers and the thermalization is
dominated by the neutrals.
In other words, except possibly in extreme ionization conditions, which may not be of
interest because the amount of {\htwoo} left is small, in general the collisions with 
electrons will not suffice to drive the {\htwoo} molecule to vibrational LTE. 
The other vibrational states  offer similar conclusions. 
According to Table \ref{ndcrit_table}, the critical densities 
for the stretching modes $v$=(100) and (001) differ by factors of a few. 
In reality, they are similar due to rapid 
interconversion collisions.
\\

The critical densities quoted in Table \ref{ndcrit_table} must be viewed as upper limits. 
Self-absorption of the photons, especially in {\htwoo}-rich atmospheres, will 
effectively reduce the probability of the upper state in the lines to radiate
thereby reducing the critical densities.
Because opacity effects are stronger in purely rotational transitions than in 
ro-vibrational transitions, the reduction will affect rotational thermalization more
severely than vibrational thermalization.\\

It is worth considering whether the departure of the {\htwoo} molecule
from LTE might be detectable with transmission spectroscopy. Focusing again on $v$=(010), which acts as the main
reservoir of vibrationally excited {\htwoo}, 
the critical densities listed in Table \ref{ndcrit_table}
 translate into atmospheric pressures of $\sim$1 and 2$\times$10$^{-2}$ dyn cm$^{-2}$ 
for the LTE-NLTE transition in {\htwo}-dominated and {\htwoo}-rich atmospheres, respectively. 
These pressures are well below the pressures of mbars ($\sim$10$^3$ dyn cm$^{-2}$) that are often quoted 
to be probed with low-resolution transmission spectroscopy, which suggests that the
detection of {\htwoo} in {NLTE} with for example JWST is hopeless. 
The technique of high-resolution spectroscopy \citep{snellen2014,lopez-moralesetal2019}, 
which probes the cores of the molecular lines,  
might be better suited to identify the departure of {\htwoo} from LTE.
Dedicated simulations are needed to confirm this possibility.

\subsection{\label{opacity_subsec}Opacity in purely rotational and ro-vibrational lines}

From Eq. \ref{kappabb_eq}, the opacity or optical thickness
of a bound-bound radiative transition in a uniform gas is approximately:
$$
\tau_{ij}= \frac{h \lambda_{ij}}{4\pi \Delta \lambda_{ij}} ( N_i B_{ij}  - N_j B_{ji}  ), 
$$
where $\Delta \lambda_{ij}$ is the line width, and $N_i$ and $N_j$ the gas columns along the line of sight
(assuming the densities $n_i$ and $n_j$ remain constant).
Taking for $\Delta \lambda_{ij}$ the full width at half maximum (fwhm) for Doppler broadening:
\begin{equation}
\tau_{ij}= \frac{1}{16 \pi \sqrt{2\ln{2}}} \sqrt{\frac{m}{kT}} A_{ji} \lambda^3_{ij}
\left(
N_i \frac{g_j}{g_i} - N_j
\right), 
\label{tauBB_eq}
\end{equation}
and therefore the line opacity depends to first order on $A_{ji}${$\lambda^3_{ij}$}. 
The data in Appendix \ref{molecularmodel_appendix} suggest that for two lines sharing the samer
lower state, the opacity of the purely rotational line can be orders
of magnitude larger than the opacity of the ro-vibrational line. 
The reverse is expected for the net radiative brackets, which behave
approximately as $p_{ji}${$\sim$}1/$\tau_{ij}$ for $\tau_{ij}${$\gg$}1. 
We have confirmed with the numerical solutions of our NLTE model that indeed the
$p_{ji}$ values are typically notably smaller for purely rotational lines than for 
ro-vibrational ones.

\subsection{Pressure range to be investigated}

Generally, the mass loss of close-in exoplanets is driven by the deposition of 
stellar extreme-ultraviolet (XUV; wavelengths shortwards of the 912-{\AA}
Lyman continuum threshold) photons. 
For photons of a given wavelength, energy deposition peaks where the optical thickness
$\tau_{\rm{XUV}}$=$\sigma_{\rm{XUV}}${$N_{\rm{XUV}}$}{$\sim$}1, 
with $\sigma_{\rm{XUV}}$ being the appropriate cross section and 
{$N_{\rm{XUV}}$} the overhead column of the XUV-absorbing molecule.
This translates into the condition
{$p_{\tau_{XUV}=1}$}{$\sigma_{\rm{XUV}}$}/{$\mu$}{$g$}{$\sim$}1, 
which shows that the $\tau_{\rm{XUV}}${$\sim$}1 level shifts to high pressures for
atmospheres of high molecular weight $\mu$ or under the effect of large 
gravitational accelerations $g$. 
For the fiducial values $\sigma_{\rm{XUV}}${$\sim$}10$^{-17}$ cm$^2$, $\mu$ in the range 2-18 amu 
and $g$/$g_{\earth}$ in the range 1--5, it is obtained that 
$p_{\tau_{UV}=1}$ is 
3.3$\times$10$^{-4}$--1.5$\times$10$^{-2}$ dyn cm$^{-2}$. This defines approximately 
the region where most of the XUV energy is deposited and where 
the cooling by {\htwoo} might be more effective at cancelling that effect. 
Hereafter, we refer to this as the XUV-energy-deposition region.
\\

\section{\label{withoutexternalirradiation_sec} Solutions. Atmospheres that are not externally irradiated}

We have run a number of {\htwoo} NLTE simulations of idealized atmospheres
to understand the main features of the 
cooling rate and its dependence on temperature and gas composition. 
For the simulations described in this section, 
the atmosphere is not 
irradiated from the outside and the only radiation that occurs is that arising internally. 
This condition is relaxed for the simulations in {\S}\ref{withexternalirradiation_sec}. 
To keep the exercise simple yet informative,  
we have considered the gas to be isothermal and that its pressure
drops exponentially with a scale height $H$=$k T$/$\mu g$. 
The simulation domain (the `atmosphere')
is discretized over a uniform spatial grid of coordinate $z$
such that each scale height is resolved by two bins. 
The grid spans 16 scale heights from approximately 10$^2$ to 10$^{-5}$ dyn cm$^{-2}$ in 32 bins.
We performed a few additional simulations with four bins per scale 
height. They showed differences in the net cooling rates relative to the standard
grid implementation that were typically {$<$}1 {\%} and acceptable for our purposes.
We have adopted for the gravitational acceleration the terrestrial value $g_{\earth}$=980 cm s$^{-2}$,  
and explored kinetic temperatures of the background gas of 200, 800 and 1,500 K.
The abundances of {\htwo}, {\htwoo} and $e^-$ affect the molecular weight $\mu$, 
and thus the conversion between $z$ and pressure, as well as the 
collisional-radiative properties. 
We have described the gas composition in terms of two parameters 
(assumed constant throughout the atmosphere), namely: the ratio of
densities of the neutrals [\htwoo]:[\htwo] and the fractional ionization 
[$e^-$]:([\htwo]+[\htwoo]). 
We have run separate simulations for o- and p-{\htwoo} using the canonical 
[o-{\htwoo}]:[p-{\htwoo}]=3:1.  
The ratio $\Gamma^{\rm{o,BB}}$/$\Gamma^{\rm{p,BB}}$ between their net cooling rates
lies typically between $\sim$1 and $\sim$3 due to a combination of abundance and opacity effects.
This point is further discussed in {\S}\ref{orthovspararate_section}. 
Unless noted otherwise, we focus on the cooling rate $\Gamma^{\rm{o,BB}}$
due to o-{\htwoo}. 
We have calculated the cooling rates by means of both
Eqs. \ref{gammabb_radbracket_eq} and \ref{gammabb_coll_eq} using the converged MOLPOP-CEP solutions 
and confirmed that in the vast majority of simulations
they match to within 3-4 significant digits.
In the presentation of the cooling rates, we use the units in which they are
directly calculated. This choice facilitates the comparison 
with previous work and is the natural format for implementation in 
Eq. \ref{energy_equation}. 
We note however that in applications concerned with low Mach number flows, 
as in the study of the terrestrial atmosphere \citep{lopez-puertastaylor2001,
feofilovkutepov2012}, the cooling rate is often expressed in units of K/day to
convey the expected temperature response of the atmosphere under the prescribed forcing. 

\subsection{Comparison to previous work.}

\citet{neufeldmelnick1987} calculated the net cooling rates due to purely rotational transitions
within the o-{\htwoo} ground vibrational state for a gas that is not externally irradiated
and compared them to similar work by \citet{hollenbachmckee1979}. 
We have extracted from our calculations to be discussed in {\S}\ref{NLTEsuite1_subsec} those
that assume inputs consistent with the conditions in the \citet{neufeldmelnick1987} work to 
prepare their Table 4. Both our calculations and those by \citet{neufeldmelnick1987}
are presented in Table \ref{coolingefficiencynm87_table}
in the form of cooling efficiencies $\Gamma^{\rm{o,BB}}$/[o-{\htwoo}][{\htwo}]. 
We have adopted the [o-{\htwoo}]:[\htwo] ratios indicated there.
For each of our entries in the table, the left and right quantities (separated by 
a comma) are $\Gamma^{\rm{o,BB}}$ and $\Gamma^{\rm{o,BB}}_{000}$, respectively. 
\\

The cooling efficiencies based on $\Gamma^{\rm{o,BB}}_{000}$ are more relevant for comparison
with the \citet{neufeldmelnick1987} calculations.
Both sets are acceptably consistent although they differ
by factors of a few for the higher temperatures and larger $\eta$ values explored here.
As defined in Table \ref{coolingefficiencynm87_table}, $\eta$ quantifies 
amongst other things the depth within the gas of the radiating parcel. 
Also, the level of agreement is sensitive to the adopted [o-{\htwoo}]:[\htwo] ratio. 
It is difficult to identify the reasons for the differences, but 
they are likely caused by one or more of the following possibilities:
\textit{1}) Our formulation considers various vibrational states, 
whereas \citet{neufeldmelnick1987} consider only the ground vibrational state;
\textit{2}) The implemented collisional rate coefficients are different, which should 
have a stronger impact at low densities when the states do not follow a Boltzmann distribution; 
\textit{3}) Our formulation refers to a gas with an exponential decay in density and zero velocity,
whereas \citet{neufeldmelnick1987} refer to a uniform-density column
with a large velocity gradient; furthermore, it is unclear how accurate is the
conversion between the \citet{neufeldmelnick1987} formulations for zero and large velocity gradients, 
which should affect the mapping of the calculations with $\eta$. 
\\

The comparison between our cooling efficiencies based on $\Gamma^{\rm{o,BB}}$ and $\Gamma^{\rm{o,BB}}_{000}$
highlights the importance of vibrational excitation and opacity.
Indeed, Table \ref{coolingefficiencynm87_table} shows that all three possibilities
$\Gamma^{\rm{o,BB}}${$>$},{$\approx$},{$<$}$\Gamma^{\rm{o,BB}}_{000}$ occur. 
The inequality {$>$} occurs when the net effect of the vibrationally excited states is
cooling the gas. 
In turn, {$<$} occurs when one or more of the $\Gamma^{\rm{BB}}_{v}$ take negative values, which 
indicates that some of the ro-vibrational lines absorb radiation rather than emit it. 
The latter occurs when the gas is heated by radiation emitted elsewhere (typically, 
but not necessarily, from a deeper layer) yet the gas cannot fully radiate away that energy, 
which becomes locally deposited.
The situation is intimately connected to the non-local 
nature of the problem, as described further in {\S}\ref{coolingvsabsorption_subsubsec}.

\subsection{\label{NLTEsuite1_subsec} Calculations}

Figures \ref{noext_1em4_1em2_fig}-\ref{noext_xe1em3_fig} summarize our calculations for
atmospheres that are not externally irradiated.
Each panel contains multiple curves.
Black curves show the net cooling rates $\Gamma^{\rm{o,BB}}$.
The red symbols indicate $\Gamma^{\rm{o,BB}}_v$ with
 $v$=(000) (circles), (010) (diamonds), (020) (triangles), (100) (squares), 
and (001) (stars). 
Filled and open symbols are used to distinguish rates $>$0 (cooling) and $<$0 (heating). 
For making more explicit the roles of NLTE and opacity,
the magenta curves indicate the cooling rates for three idealizations that translate 
in three variations of Eq. \ref{gammabb_radbracket_eq}. 
The `LTE-thin' idealization replaces in Eq. \ref{gammabb_radbracket_eq}
the densities obtained from the NLTE solution 
by the densities in LTE, i.e. $n_j${$\rightarrow$}$n^{\rm{LTE}}_j$, 
and the radiative brackets $p_{ji}${$\rightarrow$}1; 
the `NLTE-thin' idealization replaces only $p_{ji}${$\rightarrow$}1; 
the `LTE-no-thin' idealization replaces only $n_j${$\rightarrow$}$n^{\rm{LTE}}_j$. 
\\

We begin the discussion with the simulations for
a neutral atmosphere with [{\htwoo}]:[{\htwo}]=10$^{-4}$ in Fig. \ref{noext_1em4_1em2_fig}.
In these conditions the {\htwoo} molecule is mostly excited in collisions with {\htwo} and 
opacity effects are weaker than for higher {\htwoo} abundances. 
 The calculations for $T$=200 K are the easiest to understand. 
Rotational cooling dominates over vibrational cooling and 
$\Gamma^{\rm{o,BB}}${$\approx$}$\Gamma^{\rm{o,BB}}_{000}$ throughout the atmosphere.
For such a low temperature collisions cannot overcome the 
difference in energy from the ground to the excited vibrational states. 
The fact that at high pressure  
$\Gamma^{\rm{o,BB}}_{\rm{NLTE-thin}}${$\approx$}$\Gamma^{\rm{o,BB}}_{\rm{LTE-thin}}${$\gg$}$\Gamma^{\rm{o,BB}}$  
reveals the importance of opacity, which  attenuates the cooling rate by orders of
magnitude with respect to optically thin conditions. 
For {$p$}$\gtrsim$10$^{-3}$ dyn cm$^{-2}$ 
the rotational states within $v$=(000) are to practical effects thermalized 
and $\Gamma^{\rm{o,BB}}${$\approx$}$\Gamma^{\rm{o,BB}}_{\rm{LTE-no-thin}}$. 
At lower pressures, $\Gamma^{\rm{o,BB}}$ approaches 
but remains consistently smaller than $\Gamma^{\rm{BB}}_{\rm{NLTE-thin}}$. 
The reason is that the net radiative brackets $p_{ji}$ do not tend to 1, 
as might be expected from considerations based on the overhead {\htwoo} column. 
Rather, $p_{ji}${$<$}0 for some strong lines, a 
telltale for upwelling radiation that originates from the deep layers and
becomes deposited at low pressures after traversing a sufficiently large gas column.
The increased cooling at $p${$\approx$}10$^2$ dyn cm$^{-2}$  
is a general feature of all simulations and results from keeping the bottom boundary
of the simulation domain open to outgoing radiation. 
This feature does not affect the solution at lower pressures and might be 
avoided by extending the spatial grid to higher pressures than needed. 
\\

The calculations for $T$=800 and 1,500 K reveal additional properties. 
At high pressures, the cooling is dominated by ro-vibrational lines
connecting the ground and first excited vibrational states.
The reason is that
as the temperature increases the collisional excitation of the
vibrational states becomes energetically viable and vibrational cooling efficient.
Additionally (see {\S}\ref{opacity_subsec}), 
the smaller opacities of the ro-vibrational lines facilitate the escape of the emitted photons. 
In contrast, at low pressures purely rotational cooling is dominant as 
self-absorption becomes weak for all lines.
\\

Comparison of the three panels for [\htwoo]:[\htwo]=10$^{-4}$ in Fig. \ref{noext_1em4_1em2_fig} 
shows that the departure of the net cooling rate from the LTE-no-thin idealization
occurs at higher pressures for the high-temperature simulations. 
Equivalently, NLTE affects more the net cooling rate at high temperatures.
The trend is a consequence of the 
increasing contribution of vibrational cooling at high temperatures and 
the comparative difficulty of the excited vibrational states to thermalize. 
\\

The other panels in Figs. \ref{noext_1em4_1em2_fig}-\ref{noext_1ep0_1ep2_fig} extend the calculations for neutral 
atmospheres up to [\htwoo]:[\htwo]=10$^2$, in which limit the atmosphere is virtually 
pure {\htwoo}. The cooling rates can be for the most part understood 
with the arguments introduced above, and only a few complementary points are noted here.
Firstly, as the {\htwoo} abundance increases the emitted photons become easily
absorbed by other {\htwoo} molecules. 
Increased opacity sets a limit on the net cooling rate, which causes the 
progressive flattening of the $\Gamma^{\rm{o,BB}}$ curves as [\htwoo]:[\htwo] increases. 
Secondly, as the {\htwoo} abundance increases, self-collisions facilitate the 
excitation of the {\htwoo} states.
The latter effect, together with the increased availability of {\htwoo}, 
dominates over the competing effect of self-absorption and causes that 
the net cooling rate keeps increasing at even large [\htwoo]:[\htwo] ratios. 
Lastly, in the XUV-energy-deposition region, 
$\Gamma^{\rm{o,BB}}$ attains values $\gtrsim$10$^{-4}$ erg cm$^{-3}$s$^{-1}$ 
for [\htwoo]:[\htwo]{$\gtrsim$1} and $T${$\gtrsim$}800 K.
A significant contribution to it comes from $\Gamma^{\rm{o,BB}}_{010}$.
\\

The extent to which the net cooling rates predicted here can offset the stellar
heating that drives the mass loss of close-in exoplanets will depend on the specifics
of the planet-star system (e.g. atmospheric composition and stellar spectrum) and must be investigated on a case-by-case basis. 
Such an investigation is currently under way and will be presented elsewhere. 
For the time being, we note that XUV heating rates on the order of 10$^{-6}$-10$^{-5}$
erg cm$^{-3}$s$^{-1}$ have been reported for hot Jupiters and sub-Neptunes
\citep{garciamunoz2007,garciamunozetal2020}, which suggests that {\htwoo}
might contribute to the efficient cooling of upper atmospheres provided it survives in
sufficient amounts.
\\

The calculations in Fig. \ref{noext_xe1em3_fig} 
extend some of those in Figs. \ref{noext_1em4_1em2_fig}-\ref{noext_1ep0_1ep2_fig} 
by considering a fractional ionization  
of 10$^{-3}$. In principle, the enhanced collisional excitation caused by the electrons
without additional self-absorption should increase the cooling rate. 
The effect appears however subtle for the conditions explored here.
At low temperatures, the purely rotational states remain thermalized up to low pressures
without the need for electrons, and vibrational excitation remains 
energetically unfeasible. 
At low temperature therefore the electrons modify little the overall population of {\htwoo} states
and the cooling. 
As the temperature increases, rotational cooling remains dominant at low pressures but
vibrational cooling takes over at high pressures. 
The contribution of the electrons to vibrational cooling is modest
because, focusing on $v$=(010), its excitation rate coefficient 
in collisions with electrons decreases with temperature whereas the 
rate coefficients for collisions 
with {\htwo} and {\htwoo} either increase with temperature or remain essentially temperature-independent.
For example, the ratio of band rate coefficients for (de)excitation 
of $v$=(010) in collisions with electrons and {\htwo} 
is $k^{e^-}_{v'v''}$/$k^{\rm{H}_2}_{v'v''}${$\approx$}16,000 at 200 K but $\sim$540 at 1,600 K
(Appendix \ref{molecularmodel_appendix}). 
The ratios for collisions with electrons and {\htwoo}  are $\sim$380 and 
$\sim$120. Clearly, a fractional ionization of 10$^{-3}$ is insufficient to
make a huge impact on the vibrational cooling by {\htwoo}. 
This finding is likely general in that electrons, under the conditions of planetary atmospheres, 
 contribute to the cooling rate by {\htwoo} but rarely in a dominant way. 
The idea should be tested with more realistic atmospheres.
\\

The {\htwo} molecule is an inefficient coolant when it emits at IR wavelengths from its ground electronic state. 
Using the efficiencies reported by \citet{coppolaetal2019}, 
we estimate net cooling rates due to {\htwo} at $p${$\sim$}100 dyn cm$^{-2}$ of 
{$\sim$}10$^{-8}$ and 10$^{-5}$ erg cm$^{-3}$s$^{-1}$
at temperatures of 200 and 1,500 K, respectively. Because opacity effects
remain negligible for the {\htwo} IR bands, 
the cooling rate will drop in the same way as pressure for $p${$<$}100 dyn cm$^{-2}$.
Compared to our calculations for {\htwoo} in 
Figs. \ref{noext_1em4_1em2_fig}-\ref{noext_xe1em3_fig}, the net cooling rates by {\htwo}
are, even for [{\htwoo}]:[\htwo]=10$^{-4}$, lower by at least two orders of magnitude
than what our calculations show for {\htwoo}.

\subsection{Additional insight into the solutions}

\subsubsection{LTE-NLTE transition}

It is instructive to quantify the departure of the {\htwoo} states from the Boltzmann distribution
that defines LTE. 
We have produced the ratios $n_i$/$n^{\rm{LTE}}_i$ as a function of the energies $E_i$
 for three of the simulations discussed above with $T$=800 K. 
Two simulations are for neutral atmospheres with [{\htwoo}]:[\htwo]=10$^{-4}$ and 1, respectively.  
The third one has a fractional ionization of 10$^{-3}$ and [{\htwoo}]:[\htwo]=1. 
The top three rows of Fig. \ref{LTEdeparture_fig} contain a total of six panels, 
which correspond to the three simulations at pressures {$\approx$}10$^{-2}$ 
(left column) and 10$^{-3}$ (right column) dyn cm$^{-2}$. 
\\

In all cases the rotational states within $v$=(000) are the first to thermalize and
attain $n_i$/$n^{\rm{LTE}}_i${$\approx$1}, in particular the states with low excitation energy. 
Rotational thermalization within the other vibrational states is also fast and 
causes the ratios $n_i$/$n^{\rm{LTE}}_i$ to become weakly dependent on $E_i$ yet significantly {$<$1}. 
In other words, the {\htwoo} molecule remains in vibrational NLTE, 
which translates into 
the departure of the net cooling rate from the LTE-no-thin idealization
in Figs. \ref{noext_1em4_1em2_fig}-\ref{noext_xe1em3_fig}.
As expected from the combined effects of 
enhanced collisional frequency and line opacity, 
a higher pressure or the increased abundance of 
{\htwoo} or electrons drive the {\htwoo} states towards LTE.
The information contained in Fig. \ref{LTEdeparture_fig} complements the discussion
on critical densities in section {\S}\ref{ndcrit_subsec} and Table \ref{ndcrit_table}.

\subsubsection{\label{coolingvsabsorption_subsubsec} Cooling vs. absorption}

Some of the calculations in Figs. \ref{noext_1em4_1em2_fig}-\ref{noext_xe1em3_fig} 
show that the $\Gamma_v^{\rm{o,BB}}$ become negative (open symbols) at low pressures, 
and therefore associated with heating,
 a result that is traced to the net radiative brackets becoming locally negative  
for some strong lines. The top left panel of Fig.
\ref{paneltruncated_fig} offers further insight into this phenomenon 
for a neutral atmosphere with [\htwo]:[\htwoo]=1 and $T$=800 K. 
Specifically, 
we have run simulations in which the bottom boundary 
was progressively shifted to lower pressures. It turns out that $\Gamma_{010}^{\rm{o,BB}}$ 
(diamonds in the figure; other $\Gamma_{v}^{\rm{o,BB}}$ omitted for clarity) flips
from negative to positive when the simulation domain 
excludes the atmosphere with $p${$\gtrsim$}10$^{-3}$ dyn cm$^{-2}$. 
We argue from this that $\Gamma_v^{\rm{o,BB}}${$<$}0 is caused by
upwelling radiation that, after traversing a sufficiently large column, 
becomes deposited in a layer where collisional excitation is inefficient 
and cannot radiate away the upwelling photons. 
The panel shows also that although the 
$\Gamma_v^{\rm{o,BB}}$ vary between simulations, the 
net cooling rates  seem relatively unaffected by the
size of the simulation domain.
Additional calculations presented in the top right panel support 
this interpretation.
In their preparation, we kept unchanged the density profiles but reduced the temperature in the 
lower layers with the goal of diminishing the radiation emitted from them. 
The results show that when the temperature in the lower layers becomes low enough, 
$\Gamma_{010}^{\rm{o,BB}}$ flips from negative into positive.\\

Related to the above, we have confirmed that the NLTE solutions converge to the NLTE-thin
idealization under conditions that minimize the upwelling radiation and 
its deposition. 
For a neutral atmosphere with [\htwoo]:[\htwo]=10$^{-4}$ (and therefore reduced opacity) 
and $T$=800 K, the bottom panel of 
Fig. \ref{paneltruncated_fig} shows that indeed both curves overlap when the geometrical extent of the
atmosphere is small and the high-pressure atmospheric layers are omitted (both conditions 
reducing the upwelling radiation).

\subsubsection{\label{linestatistics_subsubsec} Strong lines}

For reference, we have compiled in Table \ref{stronglines_table} some information about the strongest lines in 
a neutral atmosphere at two pressures. 
The information is organized by (decreasing) line strength or energy exchanged through the line. 
For $p${$\approx$}0.15 dyn cm$^{-2}$, most of the lines are cooling and therefore
only cooling lines are tabulated. They are 
dominated by ro-vibrational transitions with upper states in $v$=(010) or (020). 
For $p${$\approx$}3$\times$10$^{-5}$ dyn cm$^{-2}$, both cooling and heating lines 
are listed. The cooling lines involve a combination of ro-vibrational transitions 
 with upper states in $v$=(010) and purely rotational transitions within 
the ground vibrational state. 
The heating lines involve ro-vibrational transitions between $v$=(000) and (010).

\subsubsection{\label{orthovspararate_section} Isomer-specific cooling}

Figure \ref{isomercooling_fig} shows the ratio $\Gamma^{\rm{o,BB}}$/$\Gamma^{\rm{p,BB}}$ of the net cooling
rates due to o- and p-{\htwoo} for some of the simulations. 
Typically, the ratio {$\rightarrow$1} at high pressure and {$\rightarrow$3} at low
pressure, a behavior dictated by line opacity. In the 
lower layers of the atmosphere, the net radiative bracket behaves overall as the 
inverse of the density and both contributions largely cancel out in Eq. \ref{gammabb_radbracket_eq}.
In such conditions, both isomers are expected to contribute about equally, which is what is found.
In contrast, opacity effects are weak at low pressures in the atmosphere and the net
cooling rates scale approximately with the relative abundance of the isomer.
The transition between the two limits depends on the {\htwoo} abundance and temperature.
Higher [\htwoo]:[\htwo] ratios and lower temperatures enhance the significance of opacity, 
as discussed above, and this results in the ratio $\Gamma^{\rm{o,BB}}$/$\Gamma^{\rm{p,BB}}${$\sim$}1
reaching up to lower pressures.
It is not evident how to trace the transition between the two limits, which suggests that
the calculation of the total cooling rate from {\htwoo} requires separate calculations of the 
cooling rates from each isomer.

\subsubsection{\label{internalenergy_subsubsec} Internal energy of excited states}

The internal energy of the gas is calculated by summing up the internal energies 
of all the particles over all
the ionization and excitation states within each particle.
In NLTE conditions, the {\htwoo}  excitation energy 
contributes by {$\rho$}{$e_{\rm{exc}}$}=$\sum_i n_i E_i$
to the internal energy of the gas and to the total energy $\rho E$. 
We have calculated {$\rho$}{$e_{\rm{exc}}$}, 
splitting the summation into purely rotational and ro-vibrational contributions, 
and compared them to the translational energy $\rho e_{\rm{tra}}$=3{$\sum_i n_i$}{$kT$}/2. 
Figure \ref{cvratio_fig} summarizes this comparison  for some simulations. 
It appears that the purely rotational contribution 
is dominant over the vibrational contribution in the range of temperatures explored, 
regardless of whether the {\htwoo} molecule is locally in LTE or NLTE. 
The vibrational energy is by no means negligible though, especially at the higher
temperatures and pressures. 
At $p${$\gtrsim$0.01--0.1 dyn cm$^{-2}$ both the rotational and vibrational contributions
appear as plateaus characteristic of LTE, 
 which means that the total excitation energy could be
estimated from the usual statistical thermodynamics prescriptions \citep[][Chapter III]{zeldovichraizer2002}.
At lower pressures, the excitation energy should ideally 
 be based on the NLTE solutions. 
These considerations must be taken into account in the evaluation of the 
 internal energy of the gas when solving Eq. \ref{energy_equation}.

\subsubsection{\label{photodissociation_subsubsec} Photodissociation of vibrationally excited states}

Photodissociation by FUV (912--2000 {\AA}) photons, with some contribution 
from XUV photons, determine the destruction of the {\htwoo} molecule in a variety of
atmospheres.  
The existent measurements of {\htwoo} photodissociation cross sections have been obtained 
at ambient temperature, in which condition  
most of the {\htwoo} states are in the ground vibrational state. 
We are unaware of any measurements or theoretical calculations of the 
photodissociation cross sections at warm temperatures or from the excited vibrational states 
included in our molecular model. 
For the atmospheres discussed in {\S}\ref{internalenergy_subsubsec}, we have calculated the 
fraction of the {\htwoo} molecules in vibrationally excited states and shown them in Fig. \ref{xsection_fig}. 
In the NLTE region at low pressures, the population of excited vibrational states is
dominated by $v$=(010), as in the LTE prediction, yet the population is significantly below the LTE prediction. 
The  photodissociation cross section of {\htwoo} in the FUV is on the order of
a few times 10$^{-18}$ cm$^2$ \citep{hrodmarssonvandishoeck2023}. 
Based on Fig. \ref{xsection_fig}, the photodissociation cross sections
of the vibrationally excited states, in particular $v$=(010), should be about 20-80 times larger than 
the ground state cross sections to contribute comparably to the 
{\htwoo} photodissociation at 800 K. The factor drops to 5-14 at 1,500 K.
If future calculations or measurements of the photodissociation cross sections 
from the excited vibrational states prove to be this large, the relevant photodissociation
channels should be included in the formulation of the NLTE problem.

\subsubsection{\label{static_section} Static vs. dynamical atmosphere}

The acceleration of the gas in the atmosphere Doppler-shifts the emitted photons between their
emission and absorption sites, facilitating their escape once the shift becomes  
comparable to the line width. For a molecular weight $\mu$=18 amu at a fiducial temperature 
of 400 K, the Doppler fwhm of a {\htwoo} line is $\sim$1 km s$^{-1}$. Velocities of this order
are predicted for the escaping atmospheres of exoplanets, but the key question is whether 
they are reached in the XUV-energy-deposition region where {\htwoo}
contributes the most to the energy balance of the upper atmosphere. 
\\

MOLPOP-CEP does not include such a dynamical effect in the treatment of the radiative transfer equation. 
Consequently, 
the cooling rates reported here should be viewed as lower limits of what might 
occur in the atmosphere. 
Ideally, the impact of this simplification must be assessed on a case-by-case basis. 
As a first example, we have done such an assessment with the simulations of exoplanet 
Trappist-1 b presented by \citet{garciamunoz2023b}. We have calculated the true overhead
column of {\htwoo} at each point of the model spatial grid, $N$({\htwoo}). 
We have also calculated the large-velocity-gradient column $N_{\rm{lvg}}$({\htwoo}) obtained 
from equating Eqs. 4 and 5 in \citet{neufeldmelnick1987}. 
$N_{\rm{lvg}}$({\htwoo}) represents the {\htwoo} column in a virtual static atmosphere
that would result in the same attenuation of the cooling rate as in the accelerating gas.
If $N$({\htwoo}){$<$}$N_{\rm{lvg}}$({\htwoo}) the true {\htwoo} column dictates the
line opacity, whereas if $N$({\htwoo}){$>$}$N_{\rm{lvg}}$({\htwoo}) the 
Doppler shift prevails and the line opacity falls below what is predicted in a static atmosphere.
We find that for the profiles of velocity and 
{\htwoo} density predicted in the Trappist-1 b atmosphere, 
$N$({\htwoo}){$\lesssim$}$N_{\rm{lvg}}$({\htwoo}) throughout the region where {\htwoo} remains undissociated. 
In this specific example 
the Doppler-shifting of the lines should not significantly affect 
the escape of the photons and the net cooling rate with respect to the calculations
for a static atmosphere. 
\\

\section{\label{withexternalirradiation_sec}Solutions. Atmospheres that are externally irradiated}

A real atmosphere is irradiated from below by photons emerging from deeper layers and 
from above by stellar photons. 
For bound-bound radiative transitions, the molecule is initially photoexcited to an upper state. 
Table \ref{Avv_table} (Appendix \ref{molecularmodel_appendix}; last two columns) lists the rates 
at which the energy is pumped from $v$=(000) under  optically thin conditions. 
They depend on the convolved
effect of the {\htwoo} radiative properties and the stellar spectrum, and we show 
them for two stellar types.  
In both cases, most of the stellar energy that is pumped from $v$=(000) 
ends up into $v$=(001). 
Some of the excited molecules will spontaneously radiate away their energy, 
 not heating the local gas.  
Others will transfer their energy to the background gas through collisions, 
 heating it. 
Photoexcitation sets a cascade (Fig. \ref{H2O_sketch_fig}) in which the initial 
photon energy is communicated to the various {\htwoo} states through collisions and
radiation. In this section, we consider how external irradiation affects the 
net cooling rates of some of the atmospheres described above.
\\

MOLPOP-CEP can optionally consider at each boundary of the simulation domain 
isotropic radiation of radiance $\mathcal{W}${$B_{\lambda}$($T_{\rm{bb}}$)} entering the domain.
$\mathcal{W}$ is the so-called dilution factor and 
{$B_{\lambda}$($T_{\rm{bb}}$)} the blackbody radiance at temperature $T_{\rm{bb}}$. 
The flux of energy through the boundary into the simulation domain 
is {$\pi$}$\mathcal{W}${$B_{\lambda}$($T_{\rm{bb}}$)}
(units of erg cm$^{-2}$s$^{-1}$). 
At the bottom boundary, we have prescribed $\mathcal{W}$=1 and a blackbody
temperature that coincides with that of the gas, i.e. $T_{\rm{bb}}$=$T$. 
The treatment is a valid representation of an atmosphere with equilibrium temperature
$T_{\rm{eq}}${$\approx$}$T$.
For completeness, we note that MOLPOP-CEP includes the option of prescribing the external irradiation through
a user-prepared table, which might be important at short wavelengths for which the stellar
spectrum notably departs from that of a black body.
At the top boundary, we have prescribed $\mathcal{W}$=($R_{\star}$/$a$)$^2$, 
where $R_{\star}$ and $a$ are the stellar radius and planet-star orbital distance, respectively, 
and for $T_{\rm{bb}}$ a value of 2,500 K.
These choices ensure that 
the flux of energy through the top boundary is a fair representation of what the real atmosphere  
receives at IR wavelengths from a TRAPPIST-1-like star 
that subtends the solid angle {$\pi$}($R_{\star}$/$a$)$^2$. 
To specify the orbital distance, we have resorted to the definition of 
$T_{\rm{eq}}$ for a zero-albedo atmosphere without energy redistribution, for which
($R_{\star}$/$a$)$^2$=2($T_{\rm{eq}}$/{$T_{\star}$})$^4$
\citep{trauboppenheimer2010}. Therefore 
$\mathcal{W}$ depends only on the ratio between the planet equilibrium temperature and the stellar effective temperature,
for which we have adopted $T_{\rm{eq}}$=$T$ and $T_{\star}$=2,500 K.
For reference, at the top boundary $\mathcal{W}$=8.2$\times$10$^{-5}$, 0.021 and 0.26 for 
$T_{\rm{eq}}$=200, 800 and 1,500 K.
Figure \ref{solidangle_sketch_fig} sketches some of these ideas. 
\\

Figure \ref{tr1_fig} summarizes the net cooling rates for neutral atmospheres irradiated 
from above only or simultaneously from above and below. Both sets of curves are represented with the
same combination of lines (solid for net cooling; dotted for net heating) and symbols. 
They are easy to tell apart because irradiation from below reduces the net cooling 
at $p${$\gtrsim$}1-10 dyn cm$^{-2}$ but leaves the lower pressures largely unaffected. 
The figure shows for comparison the net cooling rates for atmospheres that are not 
externally irradiated (solid line, no symbols). 
For the conditions explored here, irradiation from above seems to have a larger effect on 
the net cooling rate than irradiation from below. 
The region in the atmosphere affected by stellar IR irradiation depends on the {\htwoo} abundance, 
but it appears that the condition $p${$\gtrsim$}0.1 dyn cm$^{-2}$ serves as a useful 
reference.
This pressure level is largely dictated by the condition that the 
excited vibrational states transfer their energy to the background gas through collisions, 
and it is therefore connected to the critical densities discussed 
in section {\S}\ref{ndcrit_subsec} and summarized in Table \ref{ndcrit_table}.
The higher the ratio [\htwoo]:[\htwo], 
the lower the pressure at which the energy is deposited, 
a consequence of higher opacities and deexcitation collisional frequencies. 
In general, stellar IR irradiation may reduce the net cooling rate by a factor of $\sim$2 at 
$p${$\gtrsim$}1 dyn cm$^{-2}$ with respect to the solution for atmospheres that 
are not externally irradiated, or even cause the net heating of the atmosphere in those
atmospheric layers.
\\

The last row of Fig. \ref{LTEdeparture_fig} shows the ratios $n_i$/$n^{\rm{LTE}_i}$ for a neutral 
atmosphere with [\htwoo]:[\htwo]=1 that is externally irradiated. 
The main difference with the calculations presented in the same figure for atmospheres
that are not irradiated
 is that the states 
$v$=(100) and (001) are overpopulated with respect to LTE as a result of
stellar-driven photoexcitation. 
As expected, increasing pressure drives the population towards LTE.
\\

\section{Summary and outlook}

We have built a model to investigate the effects of NLTE in the 
{\htwoo}-rich atmospheres of exoplanets. The model combines up-to-date information on the
collisional and radiative properties of {\htwoo} 
with a state-of-the-art solver for the NLTE problem. 
Using idealized atmospheres of constant temperature and pressure dropping exponentially, 
we have explored how the temperature and relative abundances of {\htwo}, {\htwoo} and $e^-$
affect the cooling through bound-bound radiative transitions
 of {\htwoo} and the LTE-NLTE transition in the atmosphere. 
The net cooling rate typically increases with increasing 
{\htwoo} abundance and, more weakly, with increasing fractional ionization. 
These factors also facilitate the
thermalization of the {\htwoo} states with respect to what occurs in a neutral {\htwo}-dominated atmosphere.
Water remains in rotational LTE up to the pressures where the stellar XUV energy is 
deposited and the atmospheric gas is accelerated to space. 
Attaining vibrational LTE requires higher densities than rotational LTE, on the order of 
$\sim$10$^{13}$ cm$^{-3}$ for $v$=(010) in a neutral {\htwo}-dominated atmosphere but 
a few times $\sim$10$^{11}$ cm$^{-3}$ in a neutral {\htwoo}-rich atmosphere. 
We have estimated the critical densities for the LTE-NLTE transition of other vibrational states, 
as the information might be useful for future searches of LTE departures 
with transmission spectroscopy.
High temperatures facilitate the collisional excitation of the {\htwoo} states and 
therefore enhance the cooling rate. As the temperature increases, the vibrational
states, typically in NLTE, contribute more significantly to the cooling. 
We emphasize the importance of {\htwoo}-{\htwoo} self-collisions and line opacity
in the NLTE treatment. 
Our simulations that include stellar IR irradiation of the atmosphere show that the 
{\htwoo} populations are notably different with respect to the case of no external irradiation,
especially for $v$=(100) and (001). 
Most of the heating caused by stellar IR photons occurs at pressures $\gtrsim$0.1 dyn cm$^{-2}$. 
This is deeper in the atmosphere than where the stellar XUV photons are deposited, but 
the exact region may depend on other conditions that need to be explored on a case-specific
basis including the stellar type and atmospheric properties. 
Looking ahead, the net cooling rates reported in this work prompt detailed investigations
of the mass loss of {\htwoo}-rich atmospheres including {\htwoo} NLTE. 
The model presented here, together with the corresponding input and output files, 
are available from the corresponding author upon reasonable request. 
The collisional and radiative data will also be made available through the EMAA database 
\href{https://dx.doi.org/10.17178/EMAA}{https://dx.doi.org/10.17178/EMAA}.

\section*{Acknowledgements}

AAR acknowledges support from the Agencia Estatal de Investigaci\'on del Ministerio
de Ciencia, Innovaci\'on y Universidades (MCIU/AEI) and the European Regional Development 
Fund (ERDF) through project PID2022-136563NB-I0.

\newpage
\clearpage

\begin{figure}[h]
    \centering
    \includegraphics[width=14.cm]{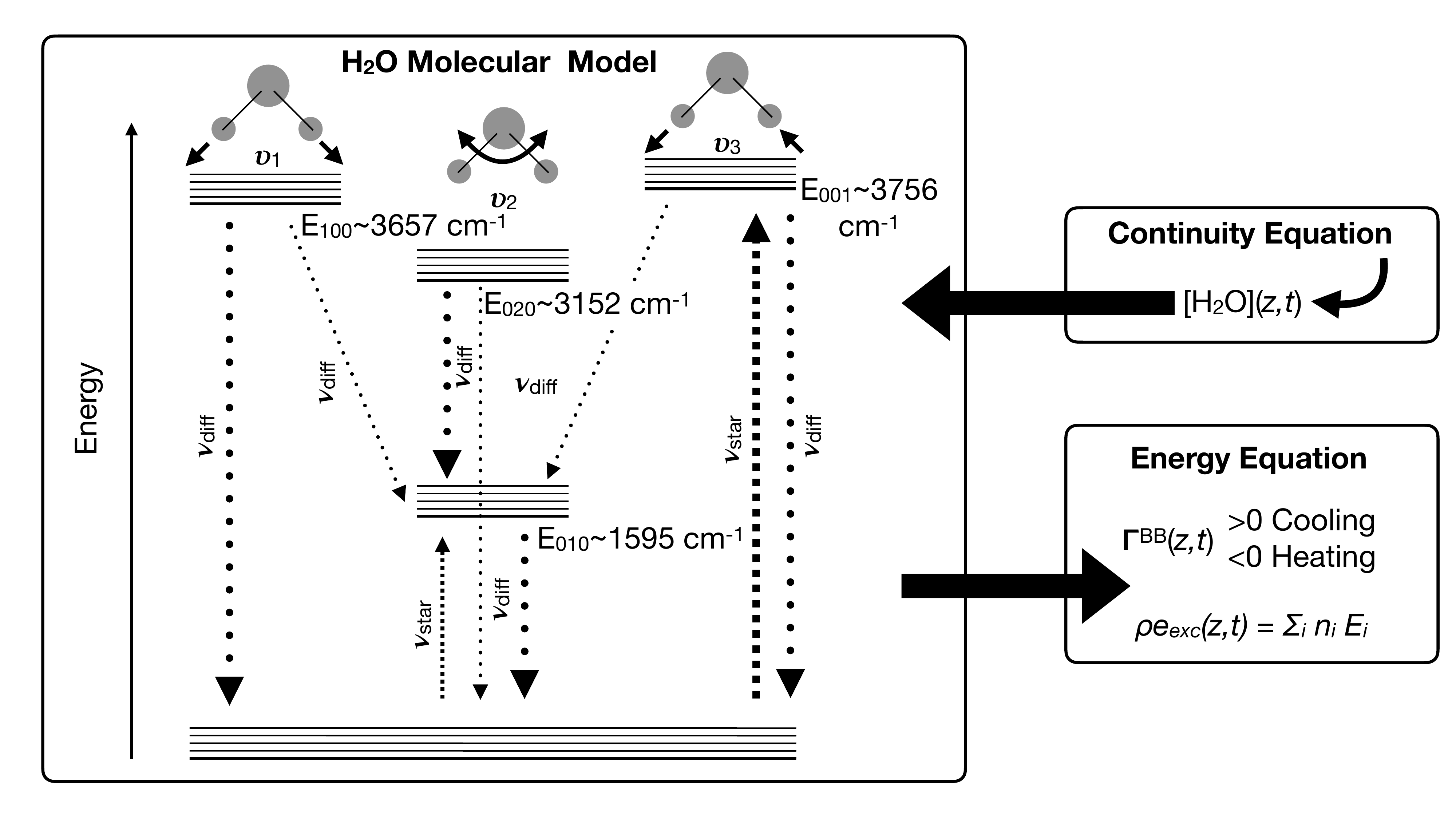}
    \caption{Sketch of our molecular model. 
    The quoted energies correspond to the lowest rotational state (including both 
    isomers) within the vibrational state. 
    For each upper vibrational state, the dotted-line arrows indicate the 
    band with the largest transition probability (big dots) and with the
    second largest transition probability (small dots) 
    (see $A_{v'v''}$ for $v'${$\neq$}$v''$, Appendix \ref{molecularmodel_appendix}). 
    For the lower vibrational state $v$=(000), the squared-line arrows indicate the photoexcitation 
    bands that carry the most of stellar energy (large squares) and the second
    most (small squares) (see {$g^{\odot}_{v''v'}$}{$hc$/$\lambda_{v'v''}$} or {$g^{\rm{tr1}}_{v''v'}$}{$hc$/$\lambda_{v'v''}$} in 
    Table \ref{Avv_table}).  
    The exchange of energy through collisions (not marked in the sketch) 
    occurs mainly from $v$=(000) into (010) with some contribution
    from $v$=(000) into (020). 
    Our treatment assumes that the total density of {\htwoo} is known at each
    location in the spatial grid $z$ and time $t$. In practice, [\htwoo] would be determined from
    solving the continuity equation of the ensamble of {\htwoo} states
     without considering a separate continuity equation for each of the individual states. 
    The outputs of the NLTE model include the net cooling rate and the     
    contribution of {\htwoo} to the internal excitation energy.
     }
    \label{H2O_sketch_fig} 
\end{figure}

\begin{figure}[h]
    \centering
    \includegraphics[width=10.5cm, angle=90]{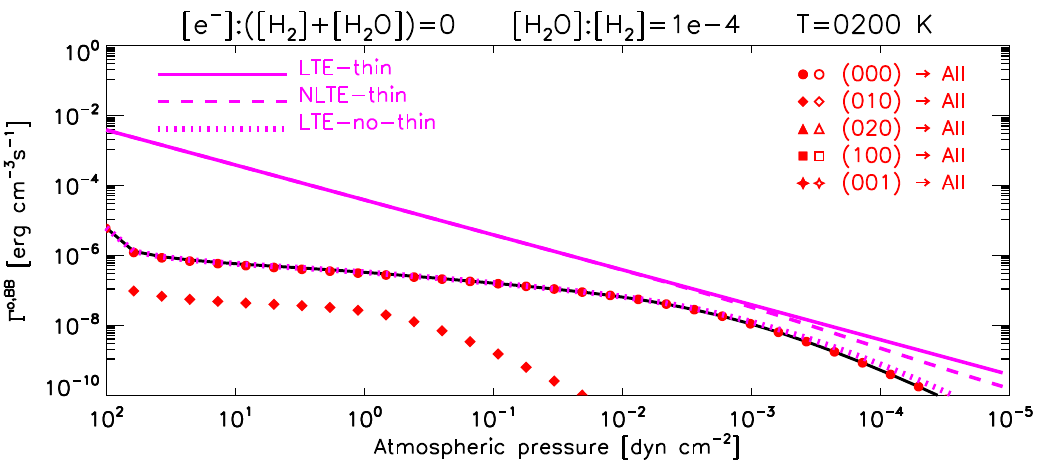}\includegraphics[width=10.5cm, angle=90]{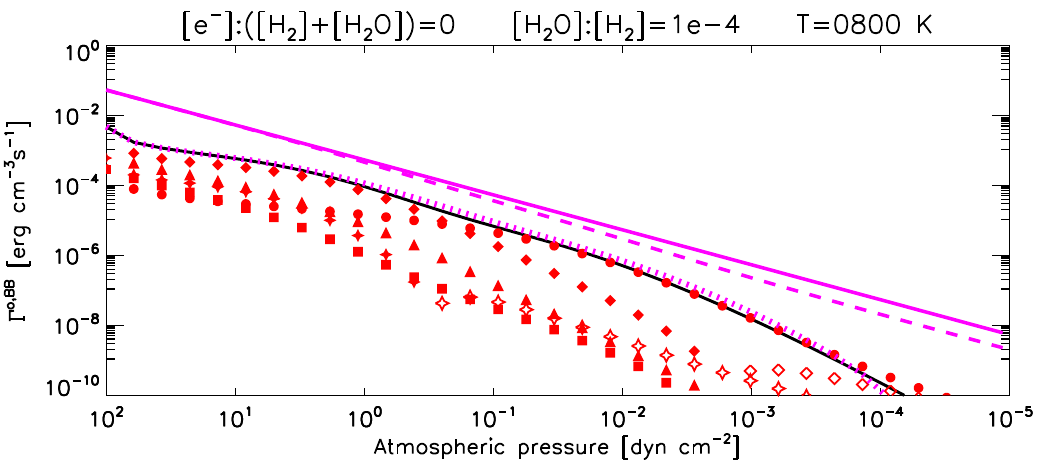}\includegraphics[width=10.5cm, angle=90]{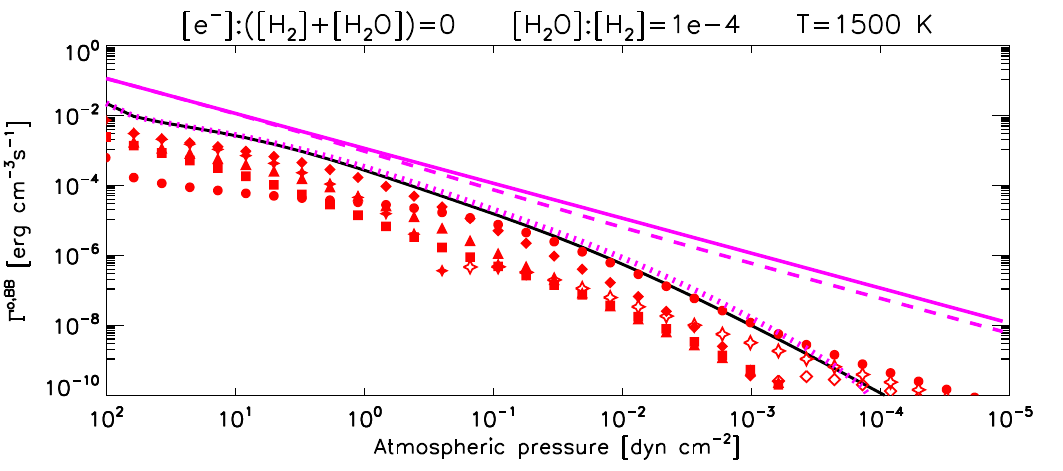}\\
    \includegraphics[width=10.5cm, angle=90]{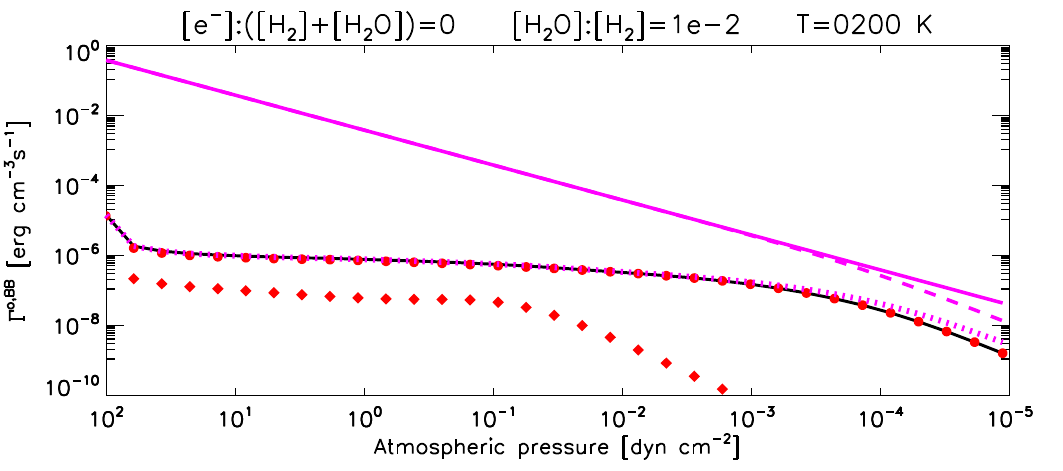}\includegraphics[width=10.5cm, angle=90]{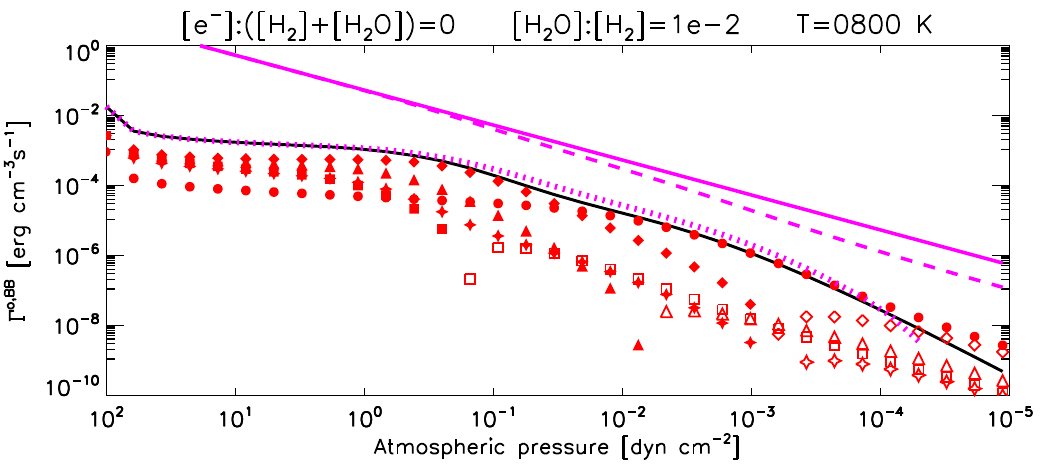}\includegraphics[width=10.5cm, angle=90]{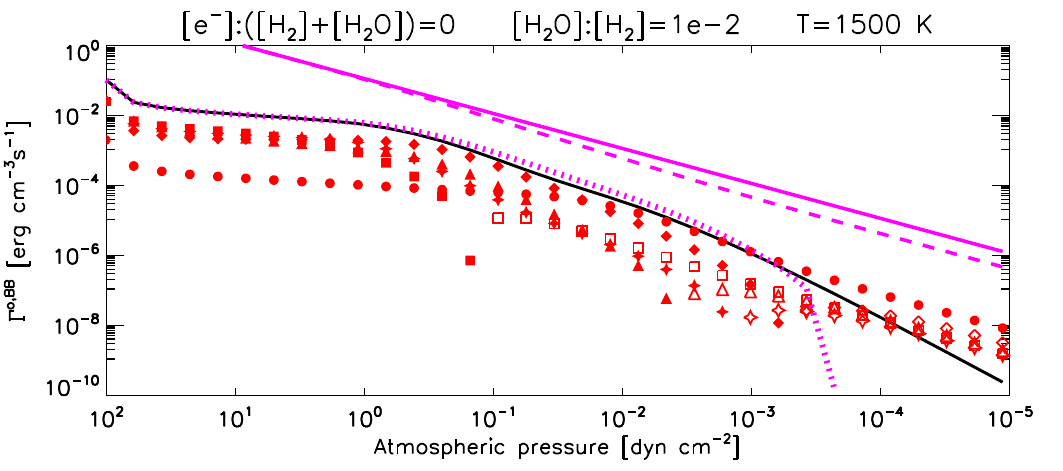}\\
    \caption{Cooling rates for atmospheres without external irradiation. 
    Conditions in panel headers. 
    Filled and open symbols refer to rates $>$0 (cooling) and $<$0 (heating), respectively. See text
    for explanation of each curve.
    }
    \label{noext_1em4_1em2_fig}
\end{figure}

\begin{figure}[h]
    \centering
    \includegraphics[width=10.5cm, angle=90]{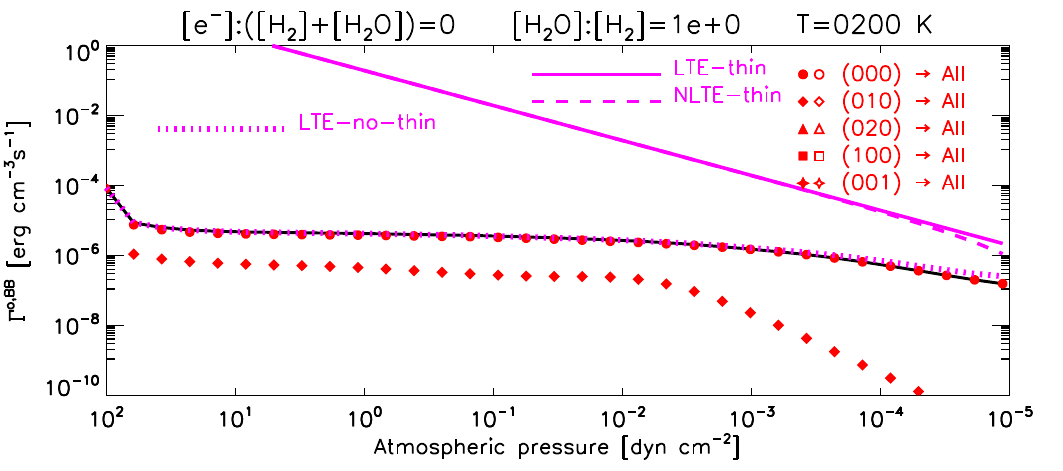}\includegraphics[width=10.5cm, angle=90]{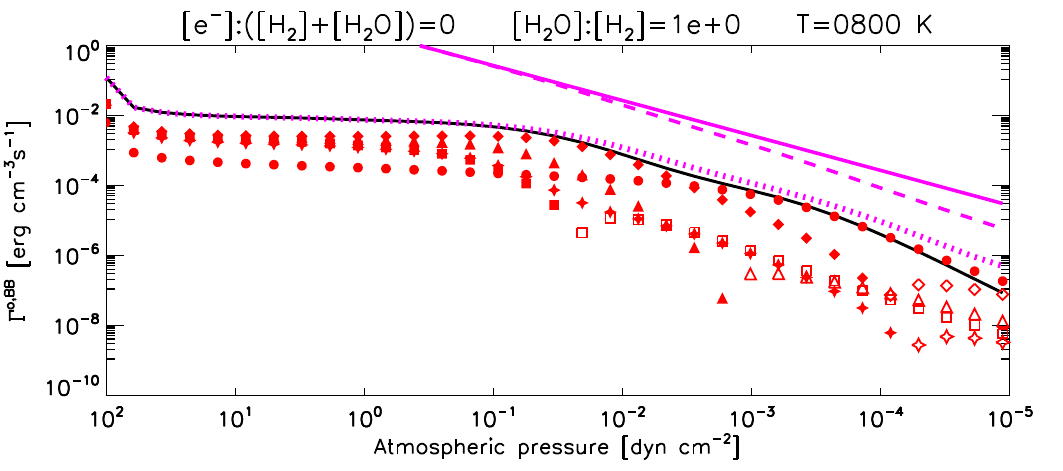}\includegraphics[width=10.5cm, angle=90]{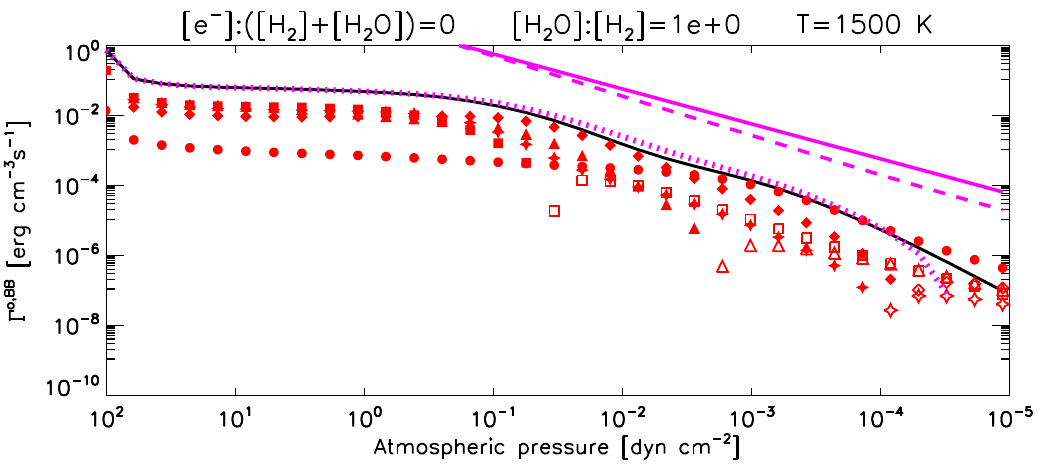}\\
    \includegraphics[width=10.5cm, angle=90]{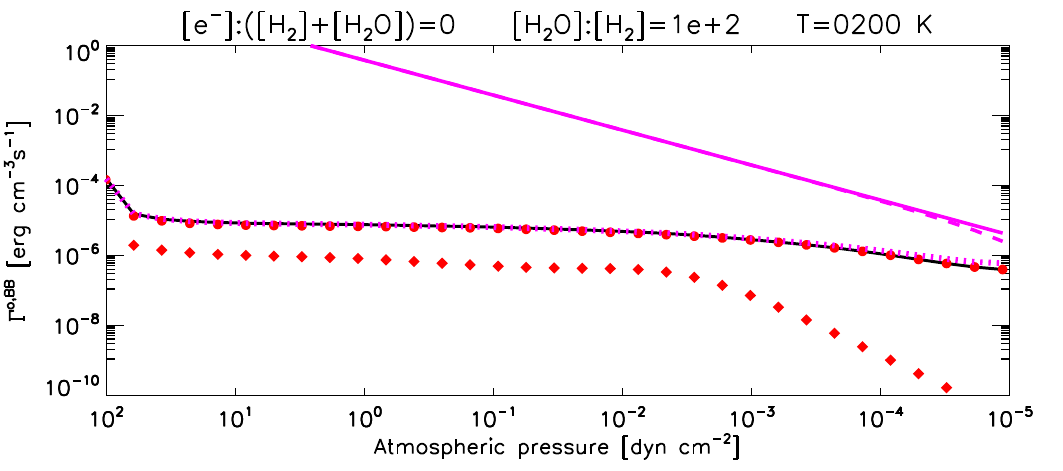}\includegraphics[width=10.5cm, angle=90]{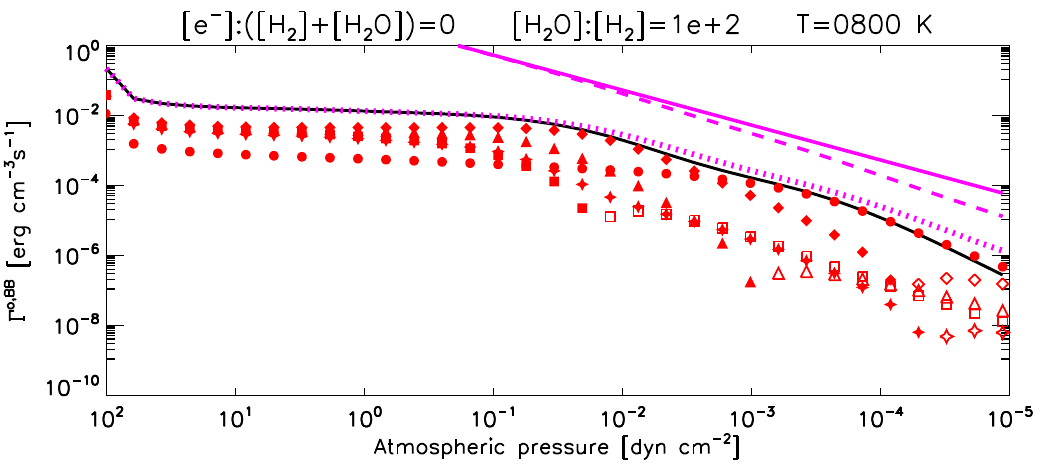}\includegraphics[width=10.5cm, angle=90]{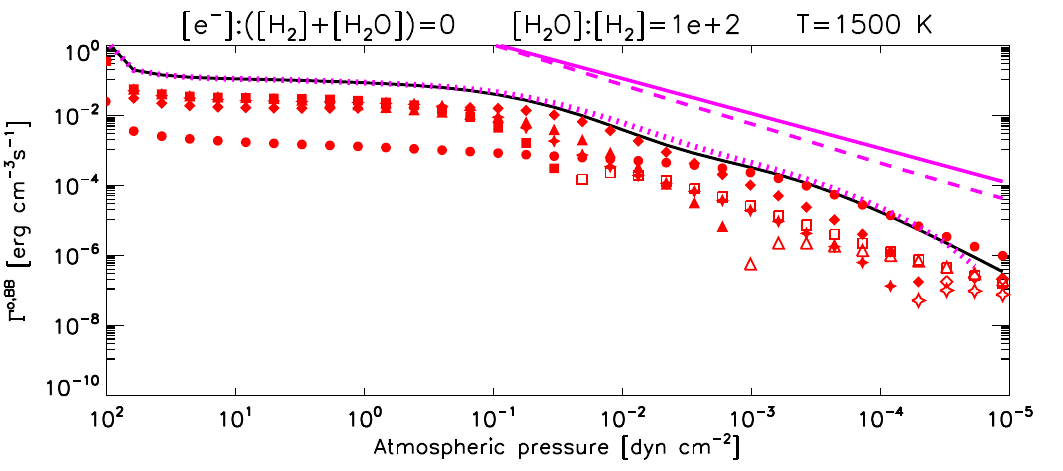}\\
    \caption{Cooling rates for atmospheres without external irradiation. 
    Conditions in panel headers.
    Filled and open symbols refer to rates $>$0 (cooling) and $<$0 (heating), respectively. See text
    for explanation of each curve.    
    }
    \label{noext_1ep0_1ep2_fig}
\end{figure}

\begin{figure}[h]
    \centering
    \includegraphics[width=10.5cm, angle=90]{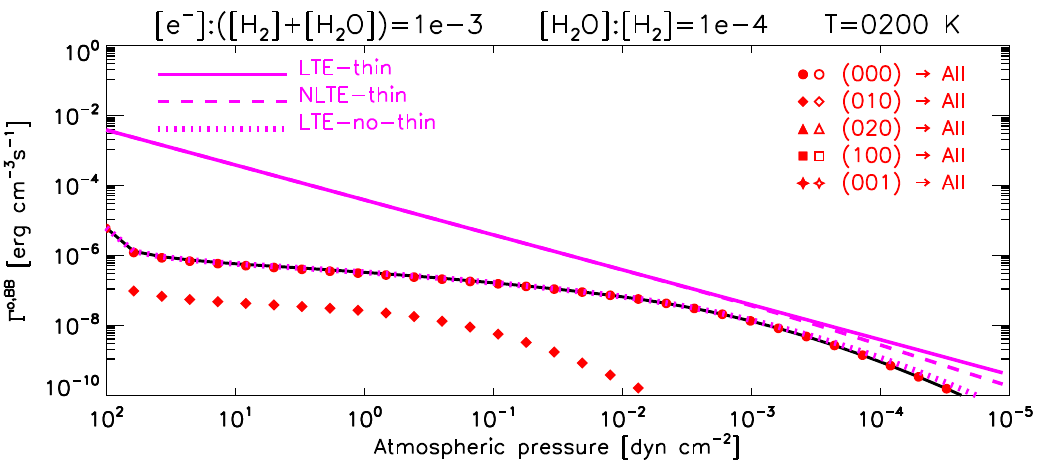}\includegraphics[width=10.5cm, angle=90]{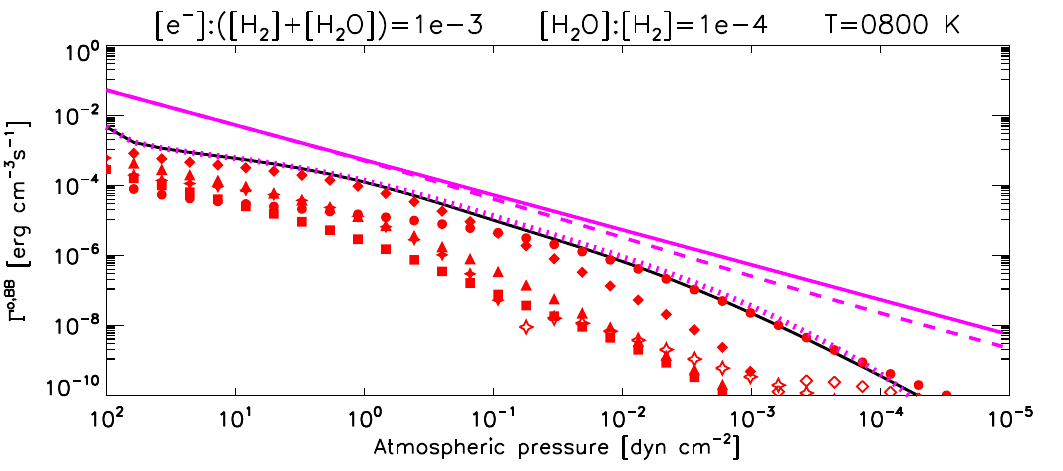}\includegraphics[width=10.5cm, angle=90]{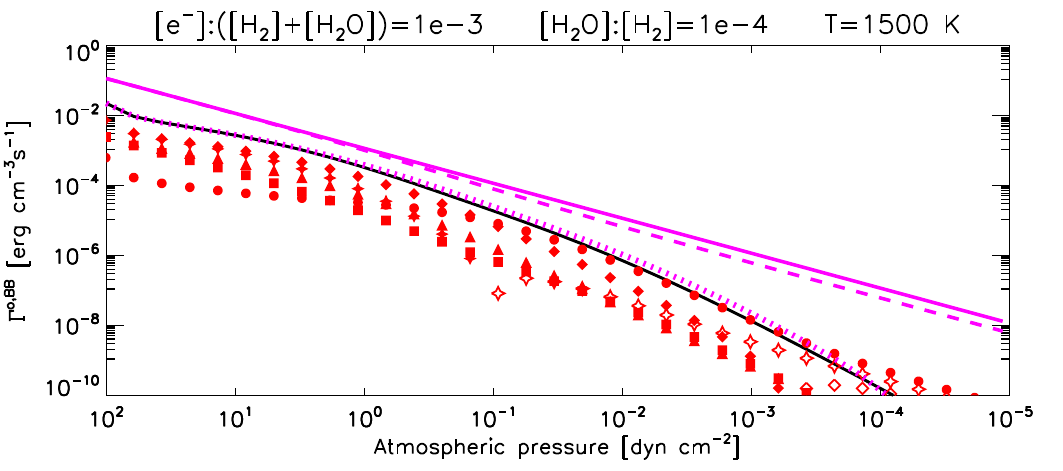}\\
    \includegraphics[width=10.5cm, angle=90]{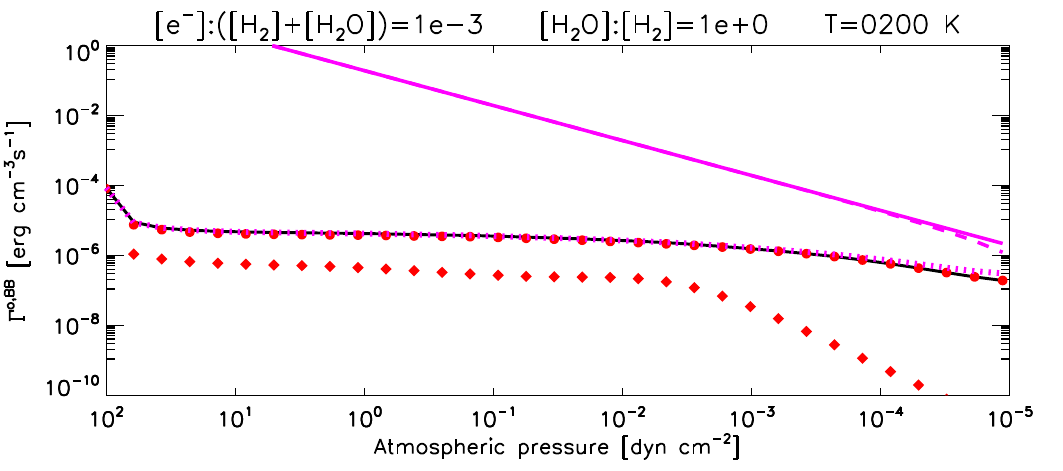}\includegraphics[width=10.5cm, angle=90]{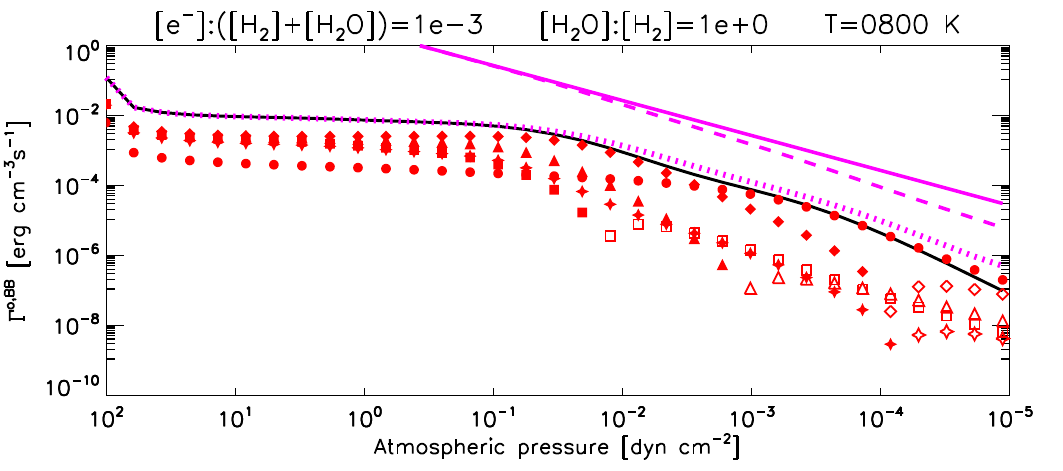}\includegraphics[width=10.5cm, angle=90]{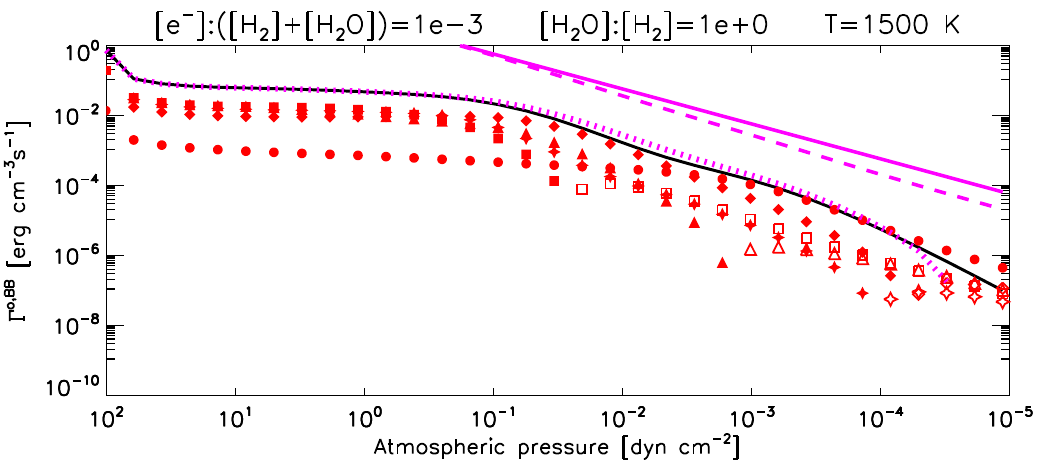}\\
    \caption{Cooling rates for atmospheres without external irradiation. 
    Conditions in panel headers.
    Filled and open symbols refer to rates $>$0 (cooling) and $<$0 (heating), respectively. See text
    for explanation of each curve.    
    }
    \label{noext_xe1em3_fig}
\end{figure}

\clearpage
\newpage

\begin{figure}[h]
    \centering
    \includegraphics[width=7cm]{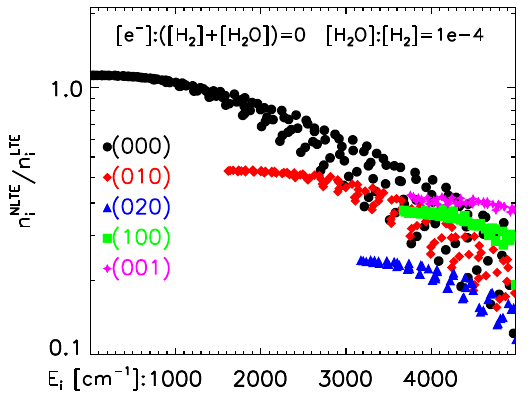}\includegraphics[width=7cm]{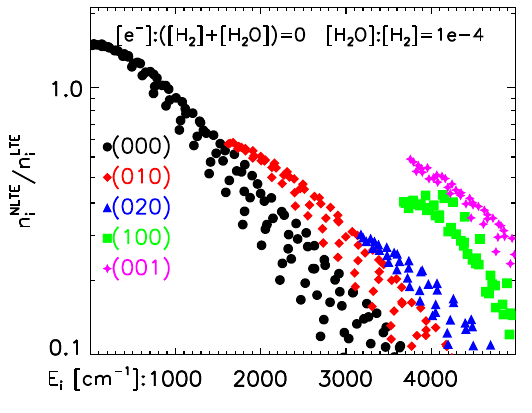}
    \vspace{-0.3cm}
    \includegraphics[width=7cm]{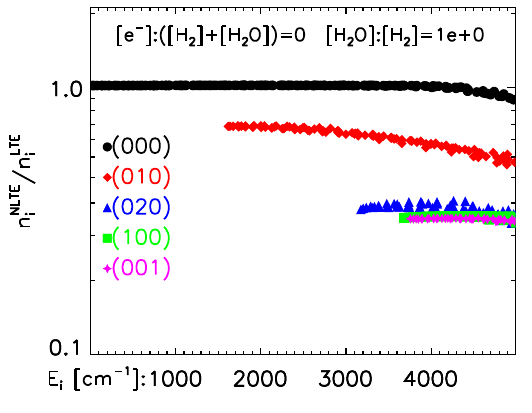}\includegraphics[width=7cm]{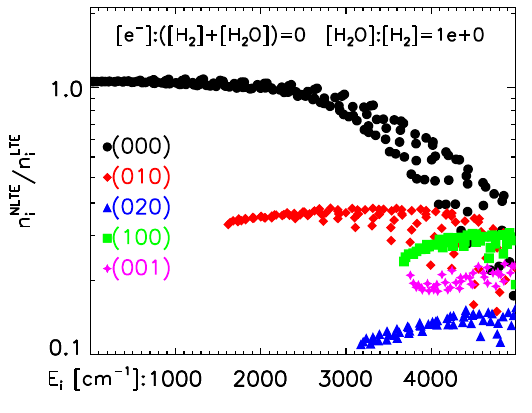}
    \vspace{-0.3cm}
    \includegraphics[width=7cm]{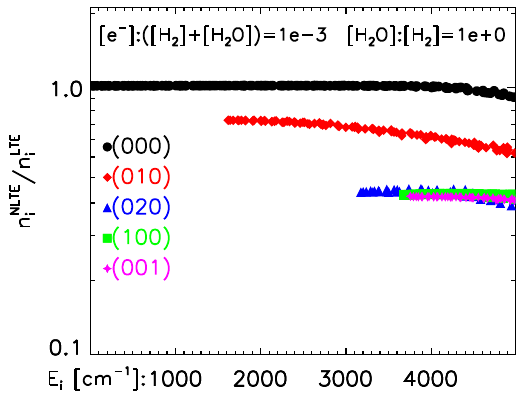}\includegraphics[width=7cm]{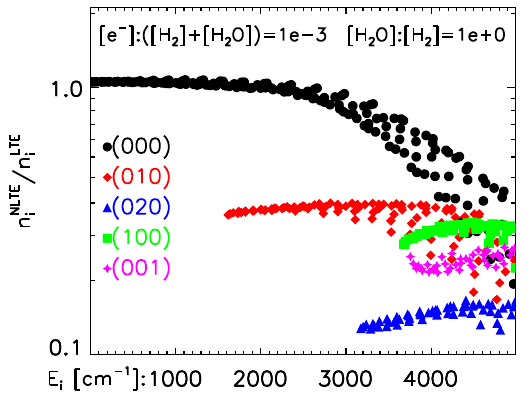}           
    \vspace{-0.3cm}
    \includegraphics[width=7cm]{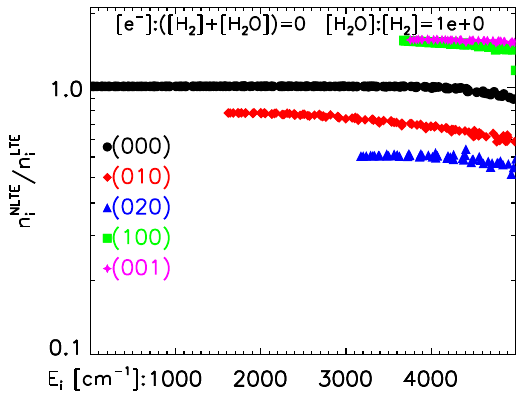}\includegraphics[width=7cm]{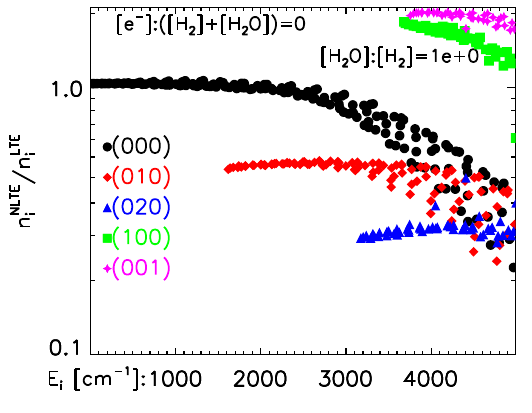}    
    \caption{Departures from LTE for the {\htwoo} states at $T$=800 K. 
    Left and right columns are for $p${$\approx$}10$^{-2}$ and 10$^{-3}$ dyn cm$^{-2}$, respectively.
    Top three rows are for atmospheres that are not externally irradiated; bottom row is
    for atmospheres that are externally irradiated. Conditions regarding fractional ionization and 
    water abundance are specified in each panel.
    }
    \label{LTEdeparture_fig}
\end{figure}

\newpage

\clearpage

\begin{figure}[h]
    \centering
    \includegraphics[width=9cm]{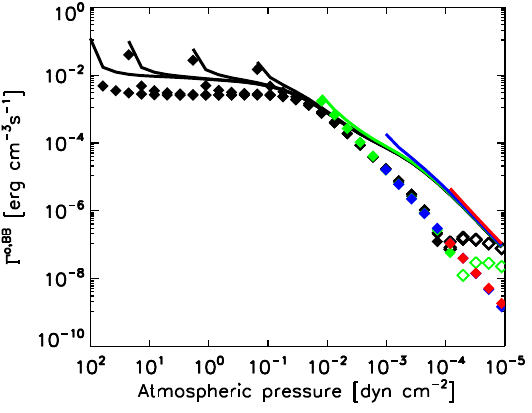}\includegraphics[width=9cm]{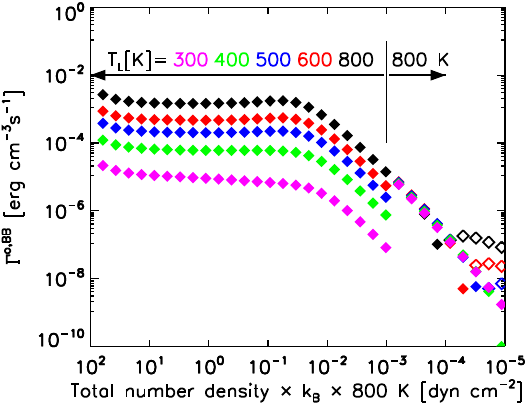}    
    \includegraphics[width=16cm]{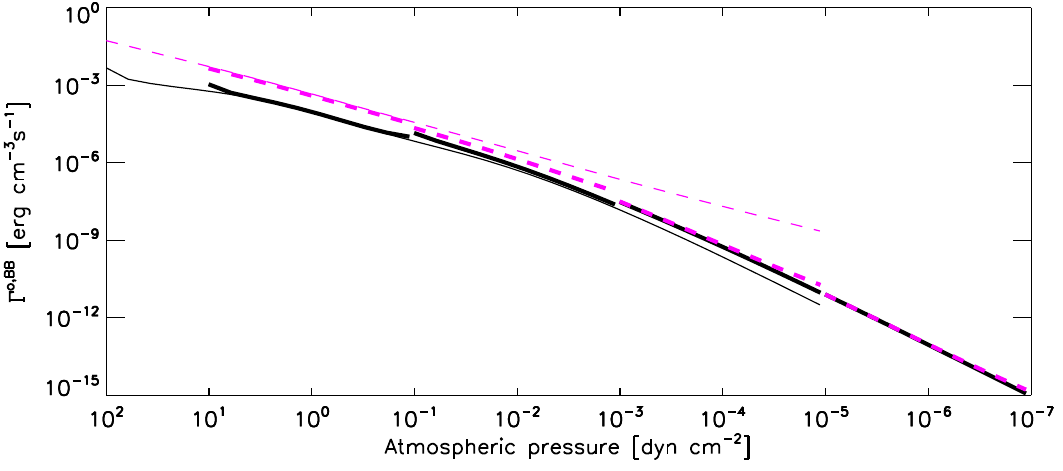}
    \caption{Top left. Net cooling rate $\Gamma^{\rm{o,BB}}$ (solid line) and partial component
    $\Gamma_{010}^{\rm{o,BB}}$ (diamonds; filled and open symbols for local cooling and heating, respectively)
    for a neutral atmosphere with [\htwoo]:[\htwo]=1 and $T$=800 K without external irradiation. 
    Colors are used to differentiate some of the calculations performed over small simulation domains.
Top right. 
Partial component of the cooling rate $\Gamma_{010}^{\rm{o,BB}}$ 
    (filled and open symbols for local cooling and heating, respectively)
    for a neutral atmosphere with [\htwoo]:[\htwo]=1 without external irradiation. 
    The NLTE calculations assume the same profiles of {\htwoo} and {\htwo} \textit{vs.} $z$
    as for the calculations of the top-left panel.    
    For the temperature, however, 
    we assumed a discontinuous two-step profile with temperature values as indicated in the figure. 
    Colors are used to differentiate the calculations for each temperature profile.
    The calculations were done with the molecular model truncated to 200 states.    
Bottom. Net cooling rate (solid black curve, thin line) for a neutral atmosphere with [\htwoo]:[\htwo]=10$^{-4}$ and $T$=800 K without external irradiation, 
as described in section \ref{NLTEsuite1_subsec}, and corresponding cooling rate
for NLTE-thin idealization (dashed magenta curve, thin line). 
Each of the black/magenta sets represented by thick lines refer to atmospheres with the
same characteristics but geometrically truncated.
    }
    \label{paneltruncated_fig}
\end{figure}

\newpage

\begin{figure}[h]
    \centering
    \includegraphics[width=9cm]{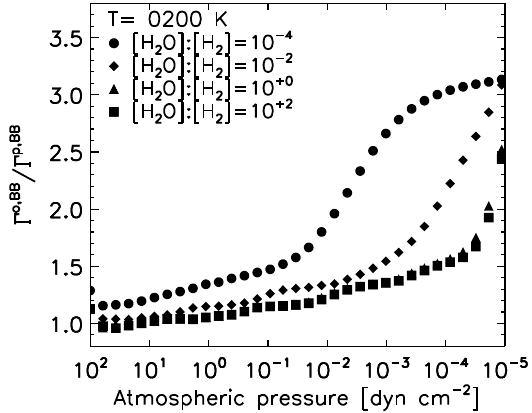}\\
    \vspace{0.4cm}
    \includegraphics[width=9cm]{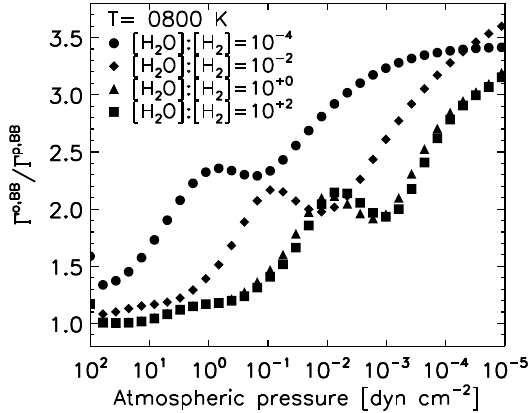}\\
    \vspace{0.4cm}    
    \includegraphics[width=9cm]{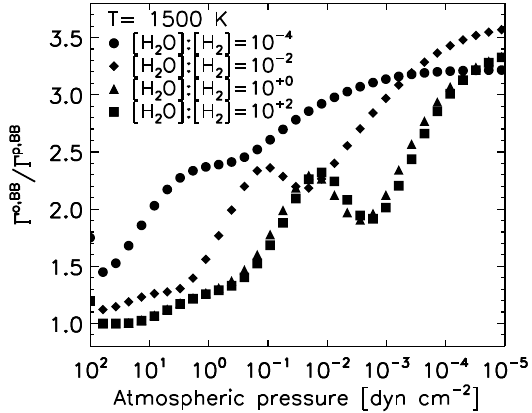}\\            
    \caption{Ratio of  net cooling rates for the two isomer forms of {\htwoo}. 
    Calculations based on the simulations of {\S}\ref{withoutexternalirradiation_sec} 
    for atmospheres that are not externally irradiated.
}
    \label{isomercooling_fig}
\end{figure}

\newpage

\begin{figure}[h]
    \centering
    \includegraphics[width=9cm]{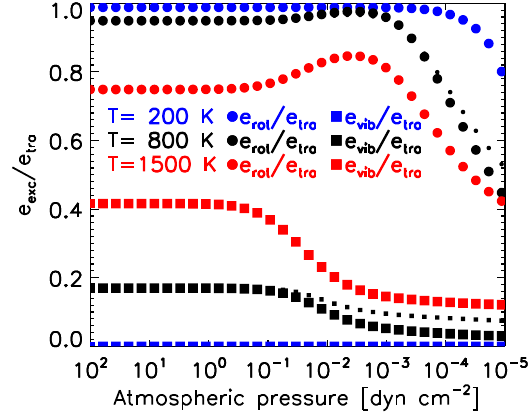}
    \caption{
    Rotational, meaning within $v$=(000), 
     and ro-vibrational components of the excitation energy
    relative to the translational energy.
    Calculations based on neutral atmospheres with [\htwoo]:[\htwo]=1 
    that are not externally irradiated. 
    Exceptionally, the small symbols are for an atmosphere of $T$=800 K that is 
    externally irradiated as described in section {\S}\ref{withexternalirradiation_sec}.
}
    \label{cvratio_fig}
\end{figure}

\begin{figure}[h]
    \centering
    \includegraphics[width=9cm]{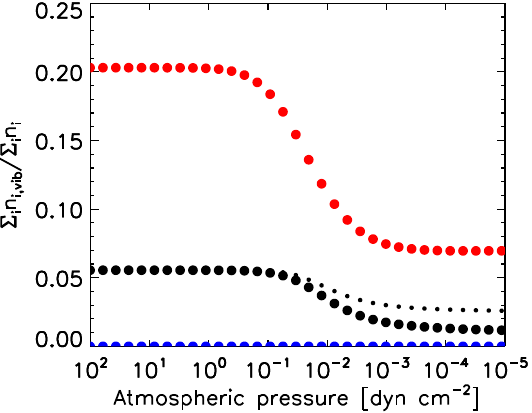}
    \caption{Fraction of excited vibrational states relative
    to the {\htwoo} total density. Same atmospheric conditions as for Fig. \ref{cvratio_fig}.
}
    \label{xsection_fig}
\end{figure}

\clearpage

\begin{figure}[h]
    \centering
    \includegraphics[width=9cm,angle=90]{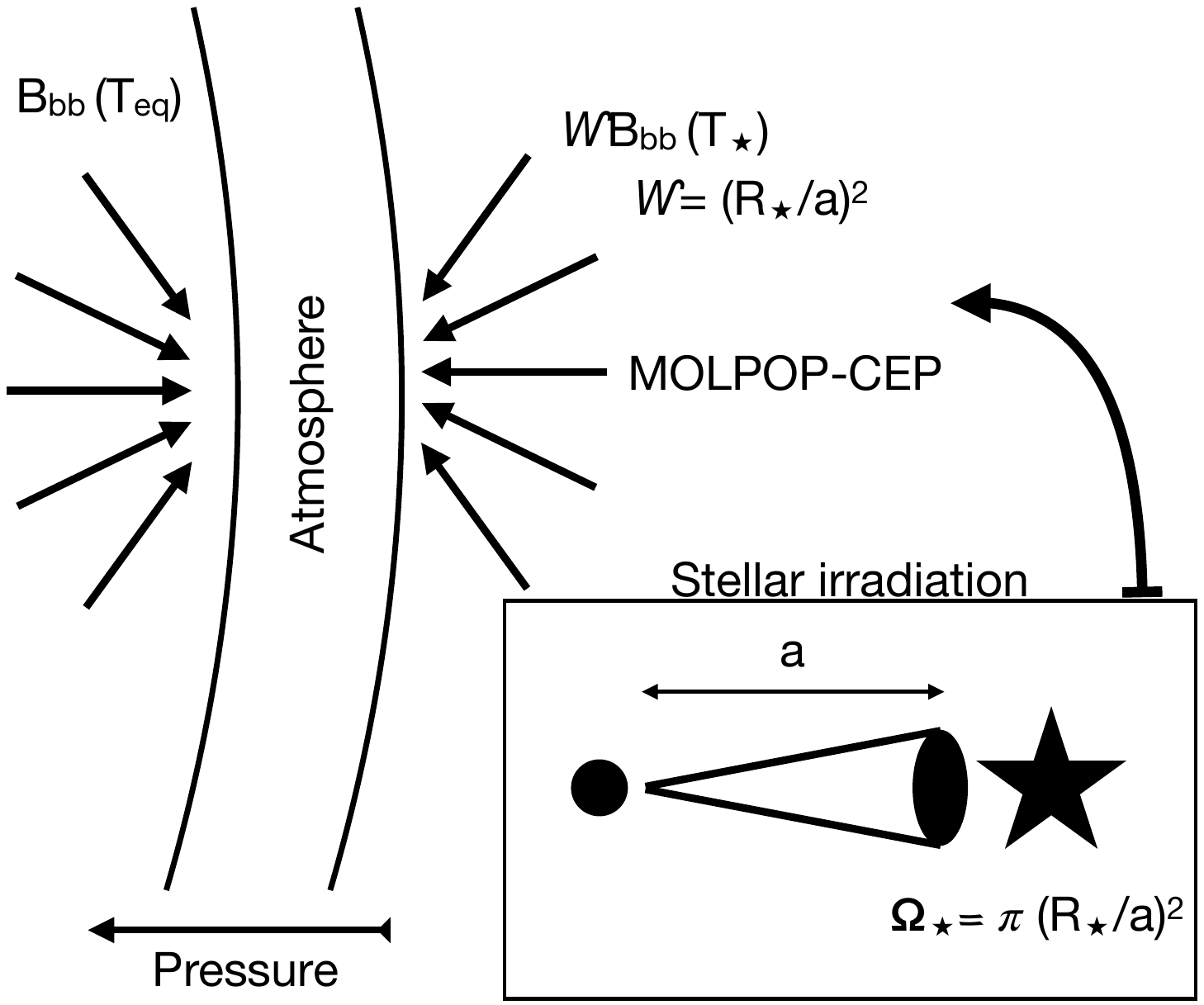}
    \caption{Simulation with MOLPOP-CEP of externally irradiated atmospheres.
    The radiation from below is implemented as isotropic blackbody radiation 
    upwelling from all directions (solid angle 2$\pi$) at the planet equilibrium temperature
    $T_{\rm{eq}}$.
    The radiation from above is implemented as isotropic blackbody radiation entering
    from all directions (solid angle 2$\pi$). The intensity of this radiation 
    is scaled by a dilution $\mathcal{W}$ to ensure that the energy 
    entering the atmosphere from above matches the energy that is received by a 
    planet of equilibrium temperature $T_{\rm{eq}}$ orbiting a star of effective temperature 
    $T_{\star}$. 
     }
    \label{solidangle_sketch_fig} 
\end{figure}

\begin{figure}[h]
    \centering
    \includegraphics[width=10.5cm, angle=90]{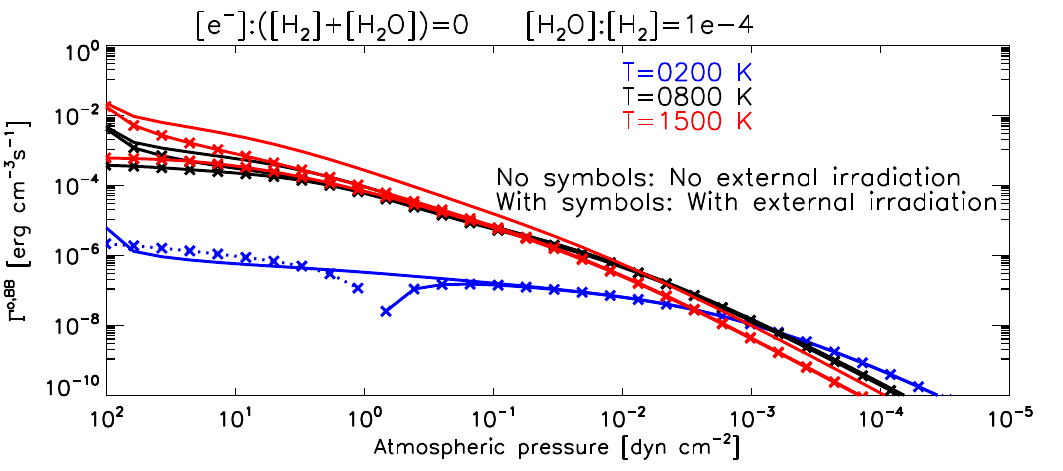}\includegraphics[width=10.5cm, angle=90]{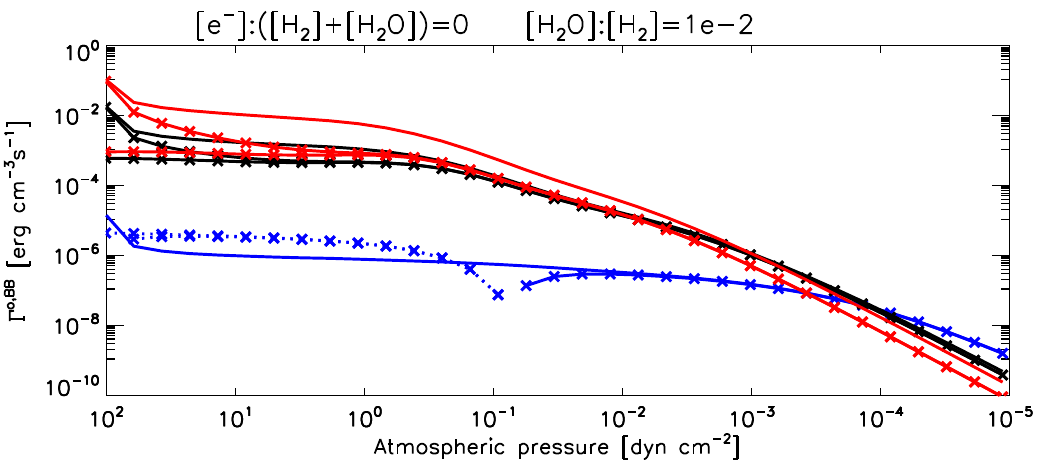}\\
    \includegraphics[width=10.5cm, angle=90]{./0_1e+0_tr1-eps-converted-to.pdf}\includegraphics[width=10.5cm, angle=90]{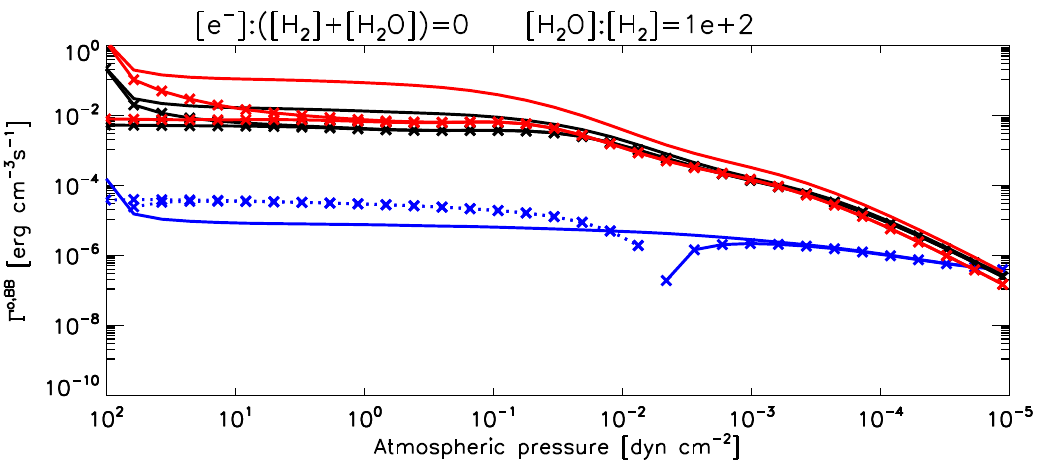}\\
    \caption{Cooling rates for atmospheres with external irradiation (as irradiated by
    a TRAPPIST 1-like star from above and a black body at the stated temperatures from below; 
    see text). 
    Conditions in panel headers. Solid and dotted lines refer to rates $>$0 (cooling)
    and $<$0 (heating), respectively.
    }
    \label{tr1_fig}
\end{figure}

\newpage
\newpage
\clearpage

\begin{table}[h]
%\begin{small}
\caption{Critical densities [cm$^{-3}$] at $T$=400 K for the vibrational thermalization of the quoted states
in collisions with one of the {\htwo}, {\htwoo} or electron particles.
Here and throughout the manuscript $X$($Y$)={$X$}{$\times$}10$^{Y}$.
}             % title of Table
\label{ndcrit_table}      % is used to refer this table in the text

\centering                          % used for centering table
\hskip-1.5cm                        
\begin{tabular}{|c| c c c c | }        % centered columns (4 columns)

\hline
                  & \multicolumn{4}{c|}{Upper vibrational state $v'$} \\
Collider          & (010) & (020) & (100) & (001) \\ 
%\cline{2-5}
\hline
{\htwo}   & 1.5($+$13) & 1.4($+$13) & 9.5($+$12) & 2.3($+$12) \\ 
{\htwoo}  & 4.3($+$11) & 4.3($+$11) & 3.7($+$11) & 7.2($+$10) \\ 
{\eminus} & 1.7($+$09) & 1.7($+$09) & 3.1($+$08) & 2.0($+$09) \\ 

\hline

\end{tabular}
%\end{small}
\end{table}

\begin{table}[h]
\begin{scriptsize}  % tiny, scriptsize, footnotesize, small
\caption{\label{coolingefficiencynm87_table}
Cooling efficiency $\Gamma^{\rm{o,BB}}$/[o-{\htwoo}][\htwo] [erg cm$^3$s$^{-1}$]
 as a function of 
{$\eta$} ({$\equiv$}$N$(o-{\htwoo})[{\htwo}]($T$/1000)$^{-0.5}${$\times$}1.35$\times$10$^{-20}$, 
cgs units; $N$ is the column to the top of the atmosphere) and  temperature. 
For each $\eta$, the top entry (in bold) refers to Table 4 of 
\citet{neufeldmelnick1987}. 
The pairs of entries immediately below are our calculations, 
arranged by the adopted [o-{\htwoo}]:[{\htwo}] ratio.
Left (regular font) and right (bold) entries are based 
on the net cooling rate $\Gamma^{\rm{o,BB}}$ and the partial component
$\Gamma^{\rm{o,BB}}_{000}$, respectively. 
Limiting $\eta${$\le$}10$^2$ helps approximately realize 
the condition established by \citet{neufeldmelnick1987} of [\htwo]{$<$}10$^{10}$
cm$^{-3}$ in their calculations.
}           
\centering
\hskip-1.5cm                        
\begin{tabular}{c c c c c c c c}

\hline
 & & \multicolumn{6}{c}{Temperature [K]} \\    % table heading 

\cline{3-8} 
$\eta$      & [o-\htwoo]:[\htwo] &  \multicolumn{1}{c}{200} &  \multicolumn{1}{c}{400} & \multicolumn{1}{c}{600} & \multicolumn{1}{c}{800} & \multicolumn{1}{c}{1000} & \multicolumn{1}{c}{1500} \\    % table heading 
\hline
            & & \bf{1.3($-$24)} & \bf{4.1($-$24)} & \bf{8.9($-$24)} & \bf{1.6($-$23)} & \bf{2.6($-$23)} & \bf{5.8($-$23)} \\

\cline{2-8} 
        &  0.75($-$5)   & 8.4($-$25),\bf{8.4($-$25)}      &   1.9($-$24),\bf{2.1($-$24)}  &  2.7($-$24),\bf{4.2($-$24)}  & 3.6($-$24),\bf{7.6($-$24)}           &   4.4($-$24),\bf{1.1($-$23)}         &    6.9($-$24),\bf{1.4($-$23)}       \\
10$^0$  &  0.75($-$4)   & 1.0($-$24),\bf{1.0($-$24)}      &   2.2($-$24),\bf{2.7($-$24)}  &  3.1($-$24),\bf{7.7($-$24)}  & 3.9($-$24),\bf{2.0($-$23)}           &   4.7($-$24),\bf{3.9($-$23)}         &    6.7($-$24),\bf{8.5($-$23)}       \\   
        &  0.75($-$3)   & 1.2($-$24),\bf{1.2($-$24)}      &   2.6($-$24),\bf{3.2($-$24)}  &  3.7($-$24),\bf{1.1($-$23)}  & 4.7($-$24),\bf{3.6($-$23)}           &   5.6($-$24),\bf{8.6($-$23)}         &    7.7($-$24),\bf{3.0($-$22)}        \\   
        &  0.75($-$2)   & 1.3($-$24),\bf{1.3($-$24)}      &   3.2($-$24),\bf{3.7($-$24)}  &  4.7($-$24),\bf{1.1($-$23)}  & 6.1($-$24),\bf{3.6($-$23)}           &   7.4($-$24),\bf{9.2($-$23)}         &    1.1($-$23),\bf{3.9($-$22)}        \\   
\hline
 & & \bf{1.1($-$24)} & \bf{3.8($-$24)} & \bf{8.4($-$24)} & \bf{1.6($-$23)} & \bf{2.5($-$23)} & \bf{5.7($-$23)} \\
\cline{2-8} 
        &  0.75($-$5)   &  6.3($-$25),\bf{6.3($-$25)}     &  1.6($-$24),\bf{1.6($-$24)}          &   2.4($-$24),\bf{2.9($-$24)}  &  3.3($-$24),\bf{4.5($-$24)}          &   4.1($-$24),\bf{5.9($-$24)}         &   6.6($-$24),\bf{8.0($-$24)}         \\
10$^1$  &  0.75($-$4)   &  8.8($-$25),\bf{8.8($-$25)}     &  2.0($-$24),\bf{2.2($-$24)}          &   2.9($-$24),\bf{4.6($-$24)}  &  3.8($-$24),\bf{9.5($-$24)}          &   4.5($-$24),\bf{1.6($-$23)}         &   6.6($-$24),\bf{3.3($-$23)}         \\   
        &  0.75($-$3)   &  1.1($-$24),\bf{1.1($-$24)}     &  2.5($-$24),\bf{2.9($-$24)}          &   3.6($-$24),\bf{8.0($-$24)}  &  4.6($-$24),\bf{2.2($-$23)}          &   5.5($-$24),\bf{4.5($-$23)}         &   7.6($-$24),\bf{1.2($-$22)}         \\   
        &  0.75($-$2)   &  1.3($-$24),\bf{1.3($-$24)}     &  3.2($-$24),\bf{3.6($-$24)}          &   4.7($-$24),\bf{1.1($-$23)}  &  6.1($-$24),\bf{3.4($-$23)}          &   7.3($-$24),\bf{8.3($-$23)}         &   1.1($-$23),\bf{3.2($-$22)}         \\   
\hline
 & & \bf{8.3($-$25)} & \bf{3.2($-$24)} & \bf{7.7($-$24)} & \bf{1.5($-$23)} & \bf{2.4($-$23)} & \bf{5.5($-$23)} \\
\cline{2-8} 
        &  0.75($-$5)   & 3.9($-$25),\bf{3.9($-$25)}      &   1.2($-$24),\bf{1.2($-$24)}         &  2.0($-$24),\bf{2.1($-$24)}  & 2.8($-$24),\bf{3.0($-$24)}       &   3.6($-$24),\bf{3.9($-$24)}         &   6.0($-$24),\bf{5.6($-$24)}         \\
10$^2$  &  0.75($-$4)   & 6.4($-$25),\bf{6.4($-$25)}      &   1.6($-$24),\bf{1.7($-$24)}         &  2.6($-$24),\bf{3.1($-$24)}  & 3.4($-$24),\bf{5.2($-$24)}       &   4.2($-$24),\bf{7.6($-$24)}         &   6.3($-$24),\bf{1.4($-$23)}         \\   
        &  0.75($-$3)   & 9.2($-$25),\bf{9.3($-$25)}      &   2.2($-$24),\bf{2.4($-$24)}         &  3.3($-$24),\bf{4.9($-$24)}  & 4.3($-$24),\bf{1.0($-$23)}       &   5.2($-$24),\bf{1.9($-$23)}         &   7.5($-$24),\bf{4.7($-$23)}         \\   
        &  0.75($-$2)   & 1.2($-$24),\bf{1.2($-$24)}      &   2.9($-$24),\bf{3.3($-$24)}         &  4.3($-$24),\bf{7.9($-$24)}  & 5.7($-$24),\bf{2.0($-$23)}       &   7.0($-$24),\bf{4.3($-$23)}         &   1.0($-$23),\bf{1.3($-$22)}         \\   

\hline

\hline
                             
\end{tabular}
\end{scriptsize}
\end{table}

\begin{table}[h]
\begin{small}
\caption{Strongest lines in a 
neutral atmosphere with [\htwoo]:[\htwo]=1 and $T$=800 K. 
I: Wavelength [$\mu$m];
II: Energy rate [erg cm$^{-3}$s$^{-1}$];
III: Transition: Upper-lower states, in the form ({$v_1$}{$v_2$}{$v_3$};$J_{\tau}$). 
For heating, the energy rate of the line is negative.
}             % title of Table
\label{stronglines_table}      % is used to refer this table in the text

\centering                          % used for centering table
\hskip-1.5cm                        
\begin{tabular}{|c| c c c | c c c | c c c |}        % centered columns (4 columns)

\hline
\multicolumn{1}{|c|}{} &
\multicolumn{3}{|c|}{$p${$\approx$}0.15 dyn cm$^{-2}$} & \multicolumn{6}{|c|}{$p${$\approx$}5$\times$10$^{-5}$ dyn cm$^{-2}$} \\
\hline
\multicolumn{1}{|c|}{} & \multicolumn{3}{|c|}{Cooling lines} & \multicolumn{3}{|c|}{Cooling lines} & \multicolumn{3}{|c|}{Heating lines} \\
\cline{2-10}
\# & I & II & III & I & II & III & I & II & III \\ 
\hline    
1 & 5.24 & 7.87($-$6) & (010;18$_{\mbox{-}17}$)--(000;17$_{\mbox{-}17}$)  &   7.04 & 5.04($-$8) & (010;3$_{3}$)--(000;4$_{3}$)                 &  5.42 &  $-$3.40($-$8) &  (010;4$_{3}$)--(000;3$_{3}$) \\ 
2 & 5.44 & 7.81($-$6) &   (020;4$_{3}$)--(010;3$_{3}$)                    &   7.17 & 4.71($-$8) & (010;4$_{1}$)--(000;5$_{3}$)                 &  5.21 &  $-$3.24($-$8) &  (010;5$_{5}$)--(000;4$_{3}$) \\ 
3 & 4.73 & 7.81($-$6) &   (010;9$_{5}$)--(000;8$_{3}$)                    &   7.21 & 4.52($-$8) & (010;4$_{3}$)--(000;5$_{5}$)                 &  5.35 &  $-$2.93($-$8) &  (010;5$_{3}$)--(000;4$_{1}$) \\ 
4 & 4.79 & 7.81($-$6) &  (010;10$_{1}$)--(000;9$_{1}$)                    &   6.86 & 4.33($-$8) & (010;2$_{1}$)--(000;3$_{3}$)                 &  5.46 &  $-$2.65($-$8) &  (010;5$_{1}$)--(000;4$_{\mbox{-}1}$) \\ 
5 & 5.11 & 7.79($-$6) & (010;15$_{\mbox{-}11}$)--(000;14$_{\mbox{-}9}$)   &   6.43 & 4.24($-$8) & (010;5$_{\mbox{-}3}$)--(000;5$_{\mbox{-}1}$) &  5.29 &  $-$2.53($-$8) &  (010;6$_{1}$)--(000;5$_{1}$) \\ 
6 & 4.68 & 7.78($-$6) &   (010;8$_{7}$)--(000;7$_{7}$)                    &   7.33 & 4.04($-$8) & (010;5$_{3}$)--(000;6$_{3}$)                 &  5.15 &  $-$2.45($-$8) &  (010;6$_{3}$)--(000;5$_{3}$) \\ 
7 & 5.17 & 7.78($-$6) & (010;16$_{\mbox{-}13}$)--(000;15$_{\mbox{-}13}$)  &  35.43 & 3.79($-$8) & (000;7$_{1}$)--(000;6$_{\mbox{-}1}$)         &  5.64 &  $-$2.34($-$8) &  (010;3$_{3}$)--(000;2$_{1}$) \\ 
8 & 5.57 & 7.78($-$6) & (020;13$_{\mbox{-}13}$)--(010;12$_{\mbox{-}11}$)  &  32.99 & 3.68($-$8) & (000;7$_{3}$)--(000;6$_{1}$)                 &  5.97 &  $-$2.33($-$8) &  (010;5$_{\mbox{-}3}$)--(000;5$_{\mbox{-}5}$) \\ 
9 & 5.36 & 7.76($-$6) &   (020;5$_{3}$)--(010;4$_{1}$)                    &  30.87 & 3.67($-$8) & (000;8$_{1}$)--(000;7$_{1}$)                 &  5.48 &  $-$2.31($-$8) &  (010;6$_{\mbox{-}1}$)--(000;5$_{\mbox{-}1}$) \\ 
10 & 5.62 & 7.75($-$6) & (020;12$_{\mbox{-}11}$)--(010;11$_{\mbox{-}11}$) &  35.94 & 3.62($-$8) & (000;6$_{3}$)--(000;5$_{3}$)                 &  5.90 &  $-$2.28($-$8) &  (010;3$_{3}$)--(000;3$_{1}$) \\ 
\hline

\end{tabular}
\end{small}
\end{table}

\newpage
\clearpage
\appendix

\section{\label{molecularmodel_appendix}Molecular model}

\subsection{\label{radiativeproperties_subsec} Energies, statistical weights and transition probabilities}

We adopted  the energies, statistical weights and transition probabilities for {\htwoo}
from the BT2 linelist  \citep{barberetal2006}, which refers to {$\sim$}28,500 
ro-vibrational states with identified quantum numbers. 
Because the cost of a NLTE calculation rises steeply with the number of states
and convergence difficulties arise when negligibly populated states 
are present, we made an effort to simplify the 
molecular model without compromising the accuracy of the NLTE calculations, as described below. %, as described below.  
%Dipole selection rules apply for radiative transitions within each of the two forms, 
%which translate into strict rules for the changes in $J$ and $\tau$ between the participating levels. 
\\

Far ultraviolet (FUV) photons excite the {\htwoo} molecule into various 
electronic states that tend to predissociate \citep{smithetal1981}. 
These photodissociation channels are generally included in the hydrodynamical models
that govern the {\htwoo} density in planetary atmospheres.  
To avoid double-counting, we omitted from our molecular model
the explicit treatment of excited electronic states. % that occur through them.
\\

Focusing on the ground electronic state, after some experimentation 
we settled on a molecular model that includes every 
ro-vibrational state with quantum numbers $v$=(000), (010), (020), (100), (001), 
$J${$\le$}20 and energies $E_i/k${$\lesssim$}7,200 K ($k$ is Boltzmann's constant). 
It is nearly identical to that used by \citet{faurejosselin2008} to describe the 
collisions of {\htwoo} with $e^-$ and {\htwo}, and includes 408 and 410 states of
o-- and p--{\htwoo} respectively.
The molecular model accommodates the majority of the radiative transitions and collisional 
channels that are significant at the temperatures and irradiation conditions of interest here. 
For reference, it includes $\sim$160 states of each isomer with up to $J$=20 in $v$=(000)
and $\sim$40 states with up to $J$=10 in $v$=(001). 
A more comprehensive molecular model would include about the same number of ro-vibrational 
states in $v$=(001) as in (000) to ensure their complete connection by stellar IR radiation.
%We explore later the uncertainties introduced by the earlier truncation of the rotational 
%states within $v$=(001) by perturbing the stellar IR radiation.
The information in Table \ref{Avv_table} confirms that the 
vibrational states $v$=(010), (020), (100) and (001) encompass $\gtrsim$90{\%} of the 
energy pumped from $v$=(000) into all the excited vibrational states for stellar radiation 
of effective temperatures 2,500 and 5,780 K. In particular, 
the energy rates associated with photoexcitation in the optically 
thin limit are shown as {$g_{v''v'}$}{$hc$/}$\lambda_{v'v''}$,
where $g_{v''v'}$ is the photoexcitation frequency. 
\\

The molecular model includes no isotopologues other than H$_2${$^{16}$}O. 
The next most abundant isotopologue in the terrestrial atmosphere is 
H$_2${$^{18}$}O and comprises 0.2{\%} of the total water \citep{galewskyetal2016}. 
Our calculations based on H$_2${$^{16}$}O (with the proper scaling of abundances)
suggest that for the {\htwoo} columns 
that we are exploring, the cooling due to H$_2${$^{18}$}O may be significant at 
high pressure, where the overhead column of {\htwoo} is large, but subdominant, 
especially at high temperatures when vibrational cooling becomes effective.

\subsection{\label{collisionalproperties_subsec}
Collisional (de)excitation.}

We are interested in {\htwo}-{\htwoo} atmospheres that, as a result of dissociation and ionization, 
may also contain electrons and other heavy particles. 
These particles undergo inelastic collisions with the target o- and p-{\htwoo}, exchanging energy with them.
We compiled up-to-date sets of state-to-state 
rate coefficients for collisions of {\htwoo} with electrons, 
{\htwo}, {\htwoo}, H, O, {\otwo} and He.
The NLTE calculations presented here focus on atmospheres composed of {\htwo}, {\htwoo} and $e^-$ only. 
Compiling the rate coefficients for collisions with the other heavy particles 
helps anticipate their significance in more complex models. 
What is meant here by electrons is the population of thermal electrons, 
assumed to have the same kinetic temperature as the heavy
particles. We defer to future work the implementation in the NLTE calculations
of non-thermal electrons. 
We discuss below the rate coefficients for deexcitation. 
MOLPOP-CEP determines those for excitation internally  
from application of detailed balancing.
\\

Setting up the NLTE model requires prescribing state-to-state rate coefficients 
$k^Q_{ji}$ for each deexcitation channel $j${$\rightarrow$}$i$ between individual 
ro-vibrational states of the o- and p-{\htwoo} target and each projectile or collider $Q$. 
This information is not always known, in which cases it must be estimated. We distinguish:
\begin{enumerate}
\item When no quantitative information on the rate coefficients exists. 
In this case, we obtained crude estimates by scaling the rate coefficients for collisions
with {\htwo} as in 
$k^Q_{ji}$=$k^{\rm{H}_2}_{ji}$($\mu_{\rm{H}_2\rm{O}-\rm{H}_2}$/$\mu_{{\rm{H}}_2{\rm{O}}-Q}$)$^{1/2}$, where 
$\mu$ (with a subscript, not to mistake for the molecular weight of the atmosphere)
stands for the reduced mass of the collision pair \citep{walkeretal2014}.
\item When the rate coefficients for rotational deexcitation 
are known for some 
states within the ground vibrational state but not for others of typically higher energy. 
\citet{neufeldmelnick1987} devised a method (hereafter, the NM87 method)
to estimate the unknown rate coefficients
from the known ones. 
It relies on the partly testable idea that the total relaxation rate coefficient is often 
weakly dependent on the upper state $j$ and may be estimated 
from the available rate coefficients, i.e. that
{$k^Q_{j}$}=$\sum_{i<j}${$k^Q_{ji}$}{$\approx$}$k_u^Q$($\downarrow$) for states $j$ above a
 threshold. 
 We adopted the NM87 method, and 
 for the state-to-state partitioning implemented their $\gamma$=0-scheme (their Eq. 6) so that 
$k^Q_{ji}$/{$k_u^Q$($\downarrow$)}=$g_i$/$\sum_{m<j}${$g_m$}.  
We estimated $k_u^Q$($\downarrow$) by averaging the total relaxation rate
coefficients {$k^Q_{j}$} over the $\sim$5--10 higher-energy 
states $j$ for which the original data report the rate coefficients. 
We mostly used the NM87 method to extrapolate, i.e. to estimate the rate coefficients
for states $j$ above the threshold used in the determination of 
{$k_u^Q$($\downarrow$)} but occasionally, in particular for {\htwoo} self-collisions, 
we used the method also for interpolation when the relevant rate coefficients were missing.
In our experience, providing relaxation channels for all the states in the molecular
model improves the numerical robustness of MOLPOP-CEP.

\item When the band rate coefficient for deexcitation between vibrational states
$k^Q_{v'v''}$ is known, but the state-to-state rate coefficients $k^Q_{ji}$ are unknown.
In this case,  
it is practical to assume that the state-to-state partitioning occurs following the 
propensity rules of collisions with another heavy particle. 
As a rule, we interpreted the band rate coefficients 
from experiments or from calculations that omit rotational details to represent thermal averages
 of the state-to-state rate coefficients.
Formally, the thermal averages are defined as $A_{v'v''}$ in Table \ref{Avv_table}, replacing $A_{ji}$ by $k^Q_{ji}$, 
and extending the summations over the relevant states of the molecular model.  
We further assumed that the branching ratios $t^Q_{ji}$=$k^Q_{ji}$/$k^Q_{v'v''}$ are 
identical to the $t^{\rm{H}_2}_{ji}$ for collisions with {\htwo}, 
which we determined from \citet{danieletal2011} and \citet{faurejosselin2008} (see below).
Given $k^Q_{v'v''}$ and $t^Q_{ji}$, the state-to-state rate coefficients $k^Q_{ji}$ are fully determined. 
By construction, the thermal averages of $k^Q_{ji}$ are equal to the prescribed $k^Q_{v'v''}$. 

\end{enumerate}

Tables \ref{kqvv_eminus_table}--\ref{kqvv_h2o_molecule_table} summarize the band rate coefficients
for collisions with electrons, {\htwo} and {\htwoo}. 
Some points worth noting from these tables and Table \ref{Avv_table} are: 
\textit{1)}
{\htwoo} radiates faster when the transition involves a change in vibrational 
state than when the transition occurs within the same vibrational state; 
\textit{2)} 
 Rotational relaxation in collisions occurs notably faster than vibrational relaxation. For example, in collisions with electrons, 
 relaxation within the ground vibrational state proceeds with a rate coefficient $\sim$5{$\times$}10$^{-7}$ cm$^3$s$^{-1}$, whereas 
 the rate coefficient for vibrational relaxation (010)$\rightarrow$(000) is $\sim$10$^{-8}$ 
 cm$^3$s$^{-1}$. 
 Based on this, the first excited vibrational state attains LTE at pressures 2-3 orders of 
 magnitude higher than required for thermalization of the rotational population;
 \textit{3)} Both rotational and vibrational relaxation of {\htwoo} occurs 2-4 orders of magnitude faster in collisions with electrons than with heavy particles. 
 As a consequence, {\htwoo} relaxation is affected by electrons once the gas fractional ionization reaches 
 {$\sim$}10$^{-4}$-10$^{-2}$, depending on temperature;
 \textit{4)} Both rotational and vibrational relaxation of {\htwoo} occur more than 
 an order of magnitude faster through self-collisions with other {\htwoo} molecules 
 than through collisions with other heavy particles. 
 Foreseeably, {\htwoo} remains in LTE up to lower pressures in a {\htwoo}-rich atmosphere 
 than in a {\htwo}-dominated atmosphere; 
 \textit{5)} The rate coefficients for collisional relaxation depend moderately on temperature. 
 Typically, the relaxation proceeds faster at lower temperatures for collisions with thermal electrons, 
 and at higher temperatures for collisions with {\htwo} or {\htwoo}.

\subsubsection{\label{collisions_electrons_sec} Collisions with electrons.}

\citet{faurejosselin2008} assembled a set of state-to-state rate coefficients for {\htwoo} 
states in collisions with thermal electrons.\footnote{The data set is currently being upgraded with new 
rate coefficients for collisions with electrons and {\htwo} 
(A. Faure, \textit{private communication}).
}
We generally adopted their rate coefficients, as downloaded from the
\href{http://cdsweb.u-strasbg.fr/cgi-bin/qcat?J/A+A/492/257}{CDS web}. 
For channels 
(010)$\rightarrow$(000), 
(020)$\rightarrow$(000), 
(020)$\rightarrow$(010), 
(100)$\rightarrow$(000),  
and
(001)$\rightarrow$(000), 
we rescaled the  data by temperature-dependent 
factors to ensure that the rescaled thermally-averaged rate coefficients match
those reported by \citet{ayouzetal2021}.
The \citet{faurejosselin2008} and \citet{ayouzetal2021} rate coefficients are generally 
consistent to within a factor of 3 except for 
(020)$\rightarrow$(000), in which case they differ by more than a factor of 10.
The band rate coefficients, calculated as thermal averages of the 
state-to-state rate coefficients, are shown in Table \ref{kqvv_eminus_table}.
 
\subsubsection{\label{collisions_h2_sec} Collisions with {\htwo}.}

\citet{phillipsetal1996}, 
\citet{faureetal2006,faureetal2007}, 
\citet{dubernetetal2006,dubernetetal2009},
\citet{danieletal2011} and
\citet{zoltowskietal2021} calculated the rate coefficients for 
rotational deexcitation within the ground vibrational state of {\htwoo}
in collisions with {\htwo} molecules. 
We adopted the \citet{danieletal2011} data for collisions with 
o- and p-{\htwo} as downloaded from 
 \href{https://home.strw.leidenuniv.nl/$\sim$moldata/H2O.html}{LAMDA}  
 \citep{schoieretal2005}
and produced rate coefficients relevant to 3:1-ortho:para Boltzmann distributions of {\htwo}. 
The \citet{danieletal2011} calculations apply to the 45 lower-energy states of 
 o- and p-{\htwoo}. 
For channels involving higher-energy states, we adopted the rate coefficients reported by
\citet{faurejosselin2008}.\\

Much less is known about the rate coefficients for channels involving excited vibrational states. 
The only relevant measurements that exist for collisions of {\htwoo} with {\htwo}
seem to be those by \citet{zittelmasturzo1991}.
\citet{faurejosselin2008} summarized the experimental and theoretical 
information available at the time and 
produced a set of state-to-state rate coefficients for collisions
involving various vibrational states. 
We adopted their deexcitation rate coefficients except for channels within the ground
vibrational state, as described above. 
Recent calculations for (010)$\rightarrow$(000)  
include those by \citet{stoecklinetal2019}, \citet{wiesenfeld2021,wiesenfeld2022} and
\citet{garcia-vazquezetal2023}, but they do not cover the temperature range investigated here.
In particular, the \citet{garcia-vazquezetal2023} work (their Fig. 9) shows that (010)
relaxation occurs with a rate coefficient $\sim$1.3$\times$10$^{-12}$ cm$^3$ s$^{-1}$ at 300 K 
that is in agreement with the experiments.
Their calculations also support the usual assumption that rotational 
relaxation occurs rapidly within each vibrational state.\\

Of importance for describing the {\htwoo} collisions with other heavy particles, 
we used the combined \citet{danieletal2011} and \citet{faurejosselin2008} 
rate coefficients to define branching ratios 
$t^{\rm{H_2}}_{ji}${$\equiv$}$k^{\rm{H_2}}_{ji}$/{$k_{v'v''}^{\rm{H_2}}$}, 
where the numerator refers to the state-to-state rate coefficient between states $j$ and $i$
(in vibrational states $v'$ and $v''$, respectively) 
and the denominator refers to the band or thermally-averaged rate coefficient. 
Table \ref{kqvv_h2_molecule_table} summarizes the band rate coefficients.

\subsubsection{\label{collisions_h2o_sec} Collisions with {\htwoo}.}

\citet{buffaetal2000}, \citet{boursieretal2020} and \citet{mandalbabikov2023} have calculated 
the rate coefficients for rotational deexcitation within the ground vibrational state of 
{\htwoo} in collisions with other {\htwoo} molecules. 
The reported calculations assume that 
the target molecule sees a Boltzmann distribution of projectiles: 
\begin{equation*}
    {\rm{H_2O}}(000,J'_{\tau'}) + {\rm{H_2O}}({\rm{thermal}}) {\rightarrow} 
    {\rm{H_2O}}(000,J''_{\tau''}) + {\rm{H_2O}}({\rm{thermal}}).
\end{equation*}
This treatment 
should be adequate here as large deviations of the {\htwoo} states
from a Boltzmann distribution only occur at very low densities where the {\htwoo} abundance 
is likely too small to make a difference in the atmospheric energy budget.
In our molecular model, we adopted 
the \citet{mandalbabikov2023} rate coefficients. We complemented them 
with those by \citet{boursieretal2020}, as downloaded from 
\href{https://basecol.vamdc.eu/index.html}{BASECOL}, which span more states.
We used the NM87 method for extrapolation to higher-energy states and for 
interpolation of a few low-energy states. 
\\

The vibrational self-relaxation of {\htwoo}
has been investigated experimentally \citep{kungcenter1975,finzietal1977,zittelmasturzo1989,
zittelmasturzo1991}
and theoretically \citep{shin1993,huestis2006,braunsteinconforti2015}.
The cross sections and rate coefficients are typically much larger than those for
collisions with other heavy particles. 
There is some evidence that {\htwoo} vibrational self-relaxation
occurs through the scheme: 
\begin{eqnarray}
{\rm{H}_2\rm{O}} (010) + \rm{X} \rightarrow {\rm{H}_2\rm{O}} (000) + \rm{X} \nonumber\\
{\rm{H}_2\rm{O}} (020) + \rm{X} \rightarrow {\rm{H}_2\rm{O}} (010) + \rm{X} \label{h2o_relaxation_scheme_eq} \\
{\rm{H}_2\rm{O}} (100,001) + \rm{X} \rightarrow {\rm{H}_2\rm{O}} (020) + \rm{X} \nonumber
\end{eqnarray}
where X{$\equiv$}{\htwoo}. We implemented the above channels  
using the band rate coefficients reported by \citet{zittelmasturzo1989} and  
omitting their weak temperature dependences. 
The scheme is consistent with the idea that the two stretching modes, 
being nearly equal in energy, behave as a single one.
For (001){$\rightarrow$}(100) and for rotational relaxation within each excited vibrational state, 
we implemented the band rate coefficient for relaxation within the ground vibrational state
to ensure rapid thermalization. 
For these cases, in which the band rate coefficient is known but not the state-to-state
partitioning, we assumed that {$t^{\rm{H}_2\rm{O}}_{ji}$}=$t^{\rm{H_2}}_{ji}$.
Table \ref{kqvv_h2o_molecule_table} summarizes this.
\\

The compilation of state-to-state rate coefficients for H, O, {\otwo} and He is described 
in the Supplementary Information.

\begin{table}[h]
\begin{small}
\caption{Radiative properties for the bands connecting some of the lower-energy vibrational 
states of {\htwoo}. $\lambda_{v'v''}$ is the wavelength.  
$A_{v'v''}$=$\sum_{j \in v' ,i \in v'', E_j >E_i} g_j \exp(-E_j/kT) A_{ji}$/$\sum_{j \in v'} g_j \exp(-E_j/kT)$ 
is the transition probability, calculated assuming that the upper states 
are thermalized at 400 K. 
$g^{\odot}_{v''v'}$ and $g^{\rm{tr1}}_{v''v'}$ are the photoexcitation 
frequencies induced 
by a Sun- ($T_{\rm{bb}}$=5,780 K) and a TRAPPIST 1-like ($T_{\rm{bb}}$=2,500 K) black-body 
of the true stellar sizes radiating from 1 AU.
The photoexcitation frequencies are calculated with Eq. 3 of \citet{crovisier1984} and
assume optically thin conditions. 
The last two columns represent the energy rates associated with photoexcitation. 
The tabulated values were calculated for o-{\htwoo} states; the values for p-{\htwoo} 
states are similar. For preparing this table, we used all rotational states
with $J${$\le$}20 within each vibrational state.
}             
\label{Avv_table}

\centering                          
\begin{tabular}{c c r l l l  l l}        

$v'$ & $v''$ & \multicolumn{1}{c}{1/$\lambda_{v'v''}$} &  \multicolumn{1}{c}{$A_{v'v''}$} & \multicolumn{1}{c}{$g^{\odot}_{v''v'}$} & \multicolumn{1}{c}{$g^{\rm{tr1}}_{v''v'}$} & \multicolumn{1}{c}{$g^{\odot}_{v''v'}${$hc$/$\lambda_{v'v''}$}} & \multicolumn{1}{c}{$g^{\rm{tr1}}_{v''v'}${$hc$/$\lambda_{v'v''}$}} \\    
                       &                         & \multicolumn{1}{c}{[cm$^{-1}$]} & \multicolumn{1}{c}{[s$^{-1}$]} & \multicolumn{1}{c}{[s$^{-1}$]} & \multicolumn{1}{c}{[s$^{-1}$]} & \multicolumn{1}{c}{[erg s$^{-1}$]} & \multicolumn{1}{c}{[erg s$^{-1}$]}\\    
\hline    
 (000) & (000)  & --  &  1.487($+$0)  &  -- &   --  &   -- & -- \\
\hline    
 (010) & (010)  & --       &  1.575($+$0)  &  -- &   --  &  -- & -- \\
       & (000)  & 1594.9  &  2.388($+$1)  &  2.649($-$4) &   1.220($-$6)  &    8.393($-$17) & 3.864($-$19) \\

 \hline
 (020) & (020)  & --       &  1.671($+$0)  &  -- &   --  &   -- & -- \\
       & (010)  & 1556.8  &  4.685($+$1)  & 5.352($-$4) &   2.483($-$6)  &   1.655($-$16) & 7.678($-$19)  \\    
       & (000)  & 3151.7  &  7.002($-$1)  & 3.178($-$6) &   1.048($-$8)  &   1.989($-$18) & 6.560($-$21) \\
 
\hline
(100) & (100)  & --      & 1.320($+$0) & --          &  --         &  -- & --  \\    
      & (020)  & 505.5   & 4.905($-$2) & 1.978($-$6) & 1.116($-$8) &  1.986($-$19) & 1.121($-$21) \\    
      & (010)  & 2062.3  & 1.246($+$0) & 1.004($-$5) & 4.203($-$8) &  4.113($-$18) & 1.722($-$20) \\    
      & (000)  & 3657.2  & 7.701($+$0) & 2.804($-$5) & 8.212($-$8) &  2.037($-$17) & 5.965($-$20) \\    

\hline

(001) & (001)  & --      & 1.379($+$0) & --          & --          &  -- & --\\    
      & (100)  & 98.8    & 1.273($-$1) & 2.765($-$5) & 1.672($-$7) &  5.424($-$19) & 3.280($-$21) \\    
      & (020)  & 604.2   & 5.705($-$2) & 1.900($-$6) & 1.054($-$8) &  2.281($-$19) & 1.265($-$21) \\    
      & (010)  & 2161.0  & 1.961($+$0) & 1.488($-$5) & 6.102($-$8) &  6.387($-$18) & 2.619($-$20) \\    
      & (000)  & 3755.9  & 9.131($+$1) & 3.191($-$4) & 9.128($-$7) &  2.381($-$16) & 6.810($-$19) \\    

\hline

(030) & (030)  & --      & 1.848($+$0) & --          & --          &  -- & -- \\    
      & (001)  & 910.8   & 3.383($-$2) & 7.187($-$7) & 3.771($-$9) &  1.300($-$19) & 6.823($-$22) \\    
      & (100)  & 1009.6  & 4.857($-$2) & 9.191($-$7) & 4.736($-$9) &  1.843($-$19) & 9.498($-$22) \\    
      & (020)  & 1515.0  & 6.912($+$1) & 8.158($-$4) & 3.816($-$6) &  2.455($-$16) & 1.148($-$18) \\    
      & (010)  & 3071.9  & 1.696($+$0) & 7.985($-$6) & 2.681($-$8) &  4.872($-$18) & 1.636($-$20)  \\    
      & (000)  & 4666.7  & 8.912($-$3) & 2.195($-$8) & 5.009($-$11) & 2.035($-$20) & 4.643($-$23) \\    
      
\hline

(110) & (110)  & --      & 1.446($+$0) & --          & --          &  -- & -- \\    
      & (030)  & 568.5   & 3.117($-$2) & 1.109($-$6) & 6.188($-$9) &  1.252($-$19) & 6.988($-$22) \\    
      & (001)  & 1479.3  & 9.379($-$1) & 1.139($-$5) & 5.366($-$8) &  3.347($-$18) & 1.577($-$20) \\    
      & (100)  & 1578.1  & 2.255($+$1) & 2.534($-$4) & 1.171($-$6) &  7.944($-$17) & 3.670($-$19) \\                        
      & (020)  & 2083.5  & 2.068($+$0) & 1.645($-$5) & 6.858($-$8) &  6.809($-$18) & 2.838($-$20) \\                              
      & (010)  & 3640.3  & 6.062($+$0) & 2.222($-$5) & 6.535($-$8) &  1.607($-$17) & 4.726($-$20) \\                              
      & (000)  & 5235.2  & 6.202($-$1) & 1.251($-$6) & 2.463($-$9) &  1.301($-$18) & 2.561($-$21) \\

\hline

(011) & (011)  & --      & 1.488($+$0) & --          & --          &  -- & -- \\    
      & (110)  & 96.1    & 8.431($-$2) & 1.884($-$5) & 1.140($-$7) &  3.594($-$19) & 2.174($-$21) \\                        
      & (030)  & 664.5   & 1.886($-$2) & 5.668($-$7) & 3.110($-$9) &  7.482($-$20) & 4.105($-$22) \\                        
      & (001)  & 1575.4  & 2.113($+$1) & 2.380($-$4) & 1.100($-$6) &  7.447($-$17) & 3.442($-$19) \\                        
      & (100)  & 1674.1  & 1.280($+$0) & 1.339($-$5) & 6.069($-$8) &  4.453($-$18) & 2.018($-$20) \\                        
      & (020)  & 2179.6  & 3.690($+$0) & 2.770($-$5) & 1.131($-$7) &  1.199($-$17) & 4.899($-$20) \\                              
      & (010)  & 3736.4  & 9.916($+$1) & 3.493($-$4) & 1.004($-$6) &  2.593($-$16) & 7.451($-$19) \\                              
      & (000)  & 5331.3  & 1.850($+$1) & 3.611($-$5) & 6.931($-$8) &  3.824($-$17) & 7.340($-$20) \\                              

\hline

(040) & (040)  & --      & 2.072($+$0) & --          & --          &  -- & -- \\    
      & (011)  & 802.6   & 8.669($-$3) & 2.120($-$7) & 1.134($-$9)  &  3.379($-$20) & 1.809($-$22) \\                              
      & (110)  & 898.6   & 5.471($-$2) & 1.180($-$6) & 6.205($-$9)  &  2.106($-$19) & 1.108($-$21) \\                        
      & (030)  & 1467.1  & 9.077($+$1) & 1.113($-$3) & 5.257($-$6)  &  3.245($-$16) & 1.532($-$18) \\                        
      & (001)  & 2377.9  & 1.740($-$3) & 1.165($-$8) & 4.562($-$11) &  5.503($-$21) & 2.155($-$23) \\                        
      & (100)  & 2476.7  & 3.228($-$3) & 2.047($-$8) & 7.848($-$11) &  1.007($-$20) & 3.861($-$23) \\                        
      & (020)  & 2982.2  & 2.758($+$0) & 1.355($-$5) & 4.643($-$8)  &  8.026($-$18) & 2.751($-$20) \\                              
      & (010)  & 4539.0  & 2.808($-$2) & 7.247($-$8) & 1.708($-$10) &  6.534($-$20) & 1.540($-$22) \\                              
      & (000)  & 6133.8  & 1.558($-$3) & 2.338($-$9) & 3.614($-$12) &  2.849($-$21) & 4.404($-$24) \\                              

\hline

\end{tabular}
\end{small}
\end{table}

\newpage

\begin{table}[h]
\caption{$k^Q_{v'v''}$: Band rate coefficients 
for {\htwoo} deexcitation in collisions with thermal electrons.  
Only the o-{\htwoo} rate  coefficients are quoted; those for p-{\htwoo} are similar.
Refs.: [FA08], \citet{faurejosselin2008}; [AY21], \citet{ayouzetal2021}.
}           
\label{kqvv_eminus_table}   
\centering                        
\begin{tabular}{c c l l l l l c}    

\hline    
$v'$ & $v''$ & \multicolumn{1}{c}{200 K} &  \multicolumn{1}{c}{400 K} & \multicolumn{1}{c}{800 K} & \multicolumn{1}{c}{1200 K} & \multicolumn{1}{c}{1600 K} & Source \\    
\hline    
 000 &  000 &   5.08($-$07) &  4.97($-$07) & 4.59($-$07) & 4.23($-$07) & 3.96($-$07)   & [FA08]  \\
 
\hline    
 010 &  000 &   2.12($-$08) & 1.44($-$08) & 9.88($-$09) & 7.83($-$09) & 6.68($-$09) & [FA08],[AY21] \\ 
\hline

 010 &  010 &   5.02($-$07) &  4.94($-$07) & 4.58($-$07) & 4.24($-$07) & 3.99($-$07) & [FA08] \\
  
 \hline     
 020 &  000 &   7.57($-$10) & 5.38($-$10) & 3.77($-$10) & 3.09($-$10) & 2.67($-$10) & [FA08],[AY21] \\
\hline
 
 020 &  010 &   3.93($-$08) & 2.72($-$08) & 1.87($-$08) & 1.48($-$08) & 1.26($-$08) & [FA08],[AY21] \\

 \hline
 
 020 &  020  &  4.94($-$07) & 4.89($-$07) & 4.57($-$07) & 4.27($-$07) & 4.03($-$07) & [FA08] \\
  
\hline    
 100 &  000 &   6.94($-$09) & 5.03($-$09) & 3.70($-$09) & 3.09($-$09) & 2.74($-$09) & [FA08],[AY21] \\

\hline
 
 100 &  010 &   1.32($-$08) & 1.19($-$08) & 9.24($-$09) & 7.70($-$09) & 6.68($-$09) & [FA08] \\

\hline
 
 100 &  020 &   1.45($-$08) & 1.20($-$08) & 8.81($-$09) & 7.12($-$09) & 6.11($-$09) & [FA08] \\

\hline
 
 100 &  100 &   5.09($-$07) & 4.98($-$07) & 4.63($-$07) & 4.32($-$07) & 4.08($-$07) & [FA08] \\
  
\hline     
 001 &  000 &   6.63($-$10) & 5.90($-$10) & 5.63($-$10) & 5.50($-$10) & 5.42($-$10) & [FA08],[AY21] \\
     
\hline
 
 001 &  010 &   6.72($-$10) & 6.36($-$10) & 5.12($-$10) & 4.74($-$10) & 4.59($-$10) & [FA08] \\

\hline
 
 001 &  020  &  1.45($-$08) & 1.21($-$08) & 8.96($-$09) & 7.27($-$09) & 6.21($-$09) & [FA08] \\
 
 \hline
 001 &  100  &  4.47($-$08) & 3.38($-$08) & 2.52($-$08) & 2.14($-$08) & 1.91($-$08) & [FA08] \\
  
 \hline
 001 &  001  &  4.94($-$07) & 4.86($-$07) & 4.54($-$07) & 4.25($-$07) & 4.03($-$07) & [FA08] \\
  
\hline
                             
\end{tabular}
\end{table}

\begin{table}[b]
\caption{$k^Q_{v'v''}$: Band rate coefficients for {\htwoo} deexcitation in collisions with a 
thermalized 3:1-ortho:para population of {\htwo}.
Only the o-{\htwoo} rate coefficients are quoted; those for p-{\htwoo} are similar.
Refs.: [FA08], \citet{faurejosselin2008}; [DA11], \citet{danieletal2011}.
}           
\label{kqvv_h2_molecule_table}   
\centering                        
\begin{tabular}{c c l l l l l c}    

\hline    
$v'$ & $v''$ & \multicolumn{1}{c}{200} &  \multicolumn{1}{c}{400} & \multicolumn{1}{c}{800} & \multicolumn{1}{c}{1200} & \multicolumn{1}{c}{1600} & Source \\    

\hline    

 000 & 000 & 2.04($-$10) & 2.54($-$10) & 3.41($-$10) & 4.24($-$10) & 4.88($-$10) & [DA11], [FA08] \\  
 010 & 000 & 1.29($-$12) & 1.55($-$12) & 3.56($-$12) & 7.12($-$12) & 1.23($-$11) & [FA08] \\  
 010 &  010 &   1.77($-$10) & 2.82($-$10) & 4.13($-$10) & 4.92($-$10) & 5.51($-$10) & [FA08] \\
 020 &  000 &   5.94($-$13) & 7.06($-$13) & 1.65($-$12) & 3.35($-$12) & 5.84($-$12) & [FA08] \\ 
 020 &  010 &   2.29($-$12) & 2.74($-$12) & 6.34($-$12) & 1.28($-$11) & 2.25($-$11) & [FA08] \\  
 020 &  020 &   1.73($-$10) & 2.77($-$10) & 4.02($-$10) & 4.75($-$10) & 5.31($-$10) & [FA08] \\  
 100 &  000 &   7.99($-$14) & 9.51($-$14) & 2.24($-$13) & 4.52($-$13) & 7.97($-$13) & [FA08] \\ 
 100 &  010 &   7.99($-$14) & 9.51($-$14) & 2.24($-$13) & 4.52($-$13) & 7.97($-$13) & [FA08] \\ 
 100 &  020 &   6.34($-$13) & 7.50($-$13) & 1.72($-$12) & 3.45($-$12) & 5.98($-$12) & [FA08] \\  
 100 &  100 &   1.81($-$10) & 2.84($-$10) & 4.04($-$10) & 4.76($-$10) & 5.32($-$10) & [FA08] \\   
 001 &  000 &   7.95($-$14) & 9.58($-$14) & 2.30($-$13) & 4.71($-$13) & 8.45($-$13) & [FA08] \\
 001 &  010 &   7.95($-$14) & 9.58($-$14) & 2.30($-$13) & 4.71($-$13) & 8.45($-$13) & [FA08] \\
 001 &  020 &   6.35($-$13) & 7.64($-$13) & 1.78($-$12) & 3.58($-$12) & 6.14($-$12) & [FA08] \\ 
 001 &  100 &   2.85($-$11) & 4.05($-$11) & 6.46($-$11) & 8.70($-$11) & 1.03($-$10) & [FA08] \\
 001 &  001 &   1.89($-$10) & 2.86($-$10) & 3.92($-$10) & 4.54($-$10) & 4.99($-$10) & [FA08] \\
  
\hline
                             
\end{tabular}
\end{table}

\begin{table}[h]
\caption{$k^Q_{v'v''}$: Band rate coefficients for {\htwoo} deexcitation in 
collisions with a thermalized 3:1-ortho:para population of {\htwoo}.
Only the o-{\htwoo} rate coefficients are quoted; those for p-{\htwoo} are similar.
Refs.: [BO20], \citet{boursieretal2020}; [MA23], \citet{mandalbabikov2023}; 
[ZI89], \citet{zittelmasturzo1989}.
}           
\label{kqvv_h2o_molecule_table}   
\centering                        
\begin{tabular}{c c l l l c}    

\hline    
$v'$ & $v''$ & \multicolumn{1}{c}{200} &  \multicolumn{1}{c}{400} & \multicolumn{1}{c}{800} & Source \\    
\hline    

 000 &  000 &   1.03($-$09) & 1.28($-$09) & 1.54($-$09) & [BO20,MA23] \\ 
     &      &   1.18($-$09) & 1.42($-$09) & 1.80($-$09) & $k_u^Q$($\downarrow$)  \\ 
 010 &  000 &   5.56($-$11) & 5.49($-$11) & 5.49($-$11) &  [ZI89] \\  
 010 &  010 &   1.02($-$09) & 1.28($-$09) & 1.54($-$09) & See text \\  
 020 &  010 &   1.10($-$10) & 1.10($-$10) & 1.10($-$10) &  [ZI89] \\
  
 020 &  020 &   1.03($-$09) & 1.28($-$09) & 1.54($-$09) & See text \\
 100 &  020 &   2.40($-$11) & 2.40($-$11) & 2.40($-$11) &  [ZI89] \\
 100 &  100 &   1.03($-$09) & 1.28($-$09) & 1.54($-$09) & See text \\
 001 &  020 &   2.40($-$11) & 2.40($-$11) & 2.40($-$11) &  [ZI89] \\
 001 &  100 &   1.03($-$09) & 1.28($-$09) & 1.54($-$09) & See text \\
 001 &  001 &   1.03($-$09) & 1.28($-$09) & 1.54($-$09) & See text \\
\hline
\end{tabular}
\end{table}

\clearpage
\newpage

\newpage

\newpage

\bibliography{steam_nlte.bib}

\clearpage

\setcounter{table}{0}
\setcounter{figure}{0}

\clearpage
\newpage

\section*{\textbf{SUPPLEMENTARY INFORMATION. I. Some Notes on MOLPOP-CEP}\\}

For our work, we have downloaded and used
version 51eef94
 of MOLPOP-CEP \citep{asensioramoselitzur2018}
available on the \href{https://github.com/aasensio/molpop-cep}{site}
in Jan 2023. 
Our approach has been to make the minimum of changes needed. 
They are of two general types:
\begin{itemize}

\item We have updated the molecular model (states, collisional and radiative properties) 
of {\htwoo} as described elsewhere in this document. We are making publicly available
the corresponding data files.

\item We have modified 3 routines of MOLPOP-CEP to produce the 
outputs needed in the calculation of the net cooling rate and other ancillary information
useful to gain insight into the solutions. The specific modifications are
described in detail below for each of the routines. We are making publicly available 
the modified MOLPOP-CEP files, the input files to run the code and reproduce our
solutions and the output files. The new outputs are written out into the four 
files: `steam.lines.dat'; `steam.coolingrates.dat'; `steam.populations.dat'; `steam.transitions.dat'.

\end{itemize}

Our simulations use the Coupled Escape Probability formalism, which produces
exact solutions to the NLTE problem in a static medium.
We use the code option to deal with media of spatially-varying properties
and omit the effects of dust.
We have optionally considered the effects of external irradiation in the simulations, which
are implemented through boundary conditions on the `left' (= bottom of the atmosphere)
and `right' (= top of the atmosphere) ends of the model. 
We have set a convergence threshold defined by 
`Accuracy in solution of the equations = 1.0e-4', which means that the maximum relative
difference between two consecutive iterations for any molecular state and altitude in the atmosphere
is less than the prescribed threshold. 
This gives undue weight to the negligibly populated states, which are often responsible for 
the overall convergence rate, but we accept this treatment because it provides extra 
confidence in the convergence of the solutions.
Additional evidence that the solutions have converged comes from the fact that the cooling 
rates obtained from Eqs. \ref{gammabb_radbracket_eq} and \ref{gammabb_coll_eq} 
generally match to within 3-4 significant digits, 
which would be difficult to explain if the numerical solution was not fully converged. 
Lastly, we ran a few additional simulations at all three kinetic temperatures of 200, 800 and 
1,500 K with a convergence threshold of `1.0e-8`. The results obtained with both convergence 
thresholds matched to 3-4 significant digits. 
We see no reasons to question the convergence of the numerical solutions presented here.
\\

\subsection*{\textbf{cep$\_$molpop$\_$interface.f90}}

The routine produces one new output file:

\begin{itemize}

\item steam.transitions.dat. It reports, for each pair of upper and lower states and
location in the spatial grid, the transition probability and the deexcitation collisional
rate of the background gas.

\end{itemize}

\subsection*{\textbf{escape$\_$cep.f90}}

The routine produces one new output file:

\begin{itemize}

\item steam.lines.dat. It reports, for each radiative transition and location in the 
spatial grid, information about the lines such as: states, wavelength, 
number density of the upper states, net radiative bracket, line emission.

\end{itemize}

\subsection*{\textbf{io$\_$cep.f90}}

The routine produces two new output files:
\begin{itemize}

\item steam.populations.dat. It reports the number densities of all the {\htwoo} states
at each location in the spatial grid. Both NLTE and LTE densities are reported.

\item steam.coolingrates.dat. It reports the cooling rates at each location in the
spatial grid. Various forms of the cooling rate are reported, including 
the net cooling rate $\Gamma^{\rm{BB}}$
as calculated through both Eqs. \ref{gammabb_radbracket_eq} and \ref{gammabb_coll_eq} and
the vibrational components $\Gamma_v^{\rm{BB}}$. 

\end{itemize}

Upon exit, the routine writes out the transition probabilities $A_{v'v''}$
and rate coefficients for deexcitation between vibrational states $k^Q_{v'v''}$. 
They are calculated at each location in the spatial grid by adding the contributions 
from the states within the upper vibrational states $v'$ as if they were in LTE.

\clearpage
\newpage

\section*{\textbf{SUPPLEMENTARY INFORMATION. II. Collisions with other heavy particles}\\}

\subsubsection*{\label{collisions_h_sec} Collisions with H.}

\citet{danieletal2015} have calculated the rate coefficients for rotational 
deexcitation within the {\htwoo} ground vibrational state in collisions with the H atom.
We  adopted them from \href{https://basecol.vamdc.eu/index.html}{BASECOL} and extrapolated them to higher-energy states with the NM87 method. 
For rotational deexcitation within the excited vibrational state (010), 
we used the branching ratios 
$t^{\rm{H_2}}_{ji}$ and assumed that the band rate coefficient is 
the same as for rotational relaxation within (000).
For vibrational deexcitation (010)$\rightarrow$(000), we used the 
branching ratios $t^{\rm{H_2}}_{ji}$ and estimated the band
rate coefficient from Fig. 9 (bottom; temperatures above 100 K) of \citet{cabrera-gonzalezetal2022}.
Table \ref{kqvv_h_atom_table} summarizes this.

\subsubsection*{\label{collisions_o_sec} Collisions with O.}

To our knowledge, the only determinations of the {\htwoo} deexcitation rate coefficients
in collisions with O atoms are the measurements for states 
(100,001) by \citet{zittelmasturzo1989}, typically assumed
to relax into (020) \citep{funkeetal2012}. 
There have been various investigations of the cross sections for rotational 
excitation within the ground vibrational state and for excitation from this into other
vibrational states \citep{dunnetal1975,kolbelgin1977,johnson1986,redmonetal1986,meyerottetal1994, 
zhouetal1994,bernsteinetal1996,braunsteinconforti2013}.
These investigations have focused on high collision energies, and 
it is not clear how to derive from them rate coefficients at temperatures $<$1,600 K.
To fill the gap for the purely rotational channels: 
$$
{\rm{H}_2\rm{O}} (000) + \rm{O} \rightarrow {\rm{H}_2\rm{O}} (000) + \rm{O}
$$
$$
{\rm{H}_2\rm{O}} (010) + \rm{O} \rightarrow {\rm{H}_2\rm{O}} (010) + \rm{O}
$$
and the vibrational channels:
$$
{\rm{H}_2\rm{O}} (010) + \rm{O} \rightarrow {\rm{H}_2\rm{O}} (000) + \rm{O} 
$$
$$
{\rm{H}_2\rm{O}} (020) + \rm{O} \rightarrow {\rm{H}_2\rm{O}} (010) + \rm{O},
$$
we adopted the corresponding rate coefficients for collisions with {\htwo}
and scaled them by $(\mu_{\rm{H}_2\rm{O}-\rm{H}_2}/\mu_{\rm{H}_2\rm{O}-\rm{O}})^{1/2}${$\approx$}0.46. 
 Table \ref{kqvv_o3p_atom_table} summarizes this.

\subsubsection*{\label{collisions_o2_sec} Collisions with {\otwo}.}

To our knowledge, there are no determinations of the rate coefficients for 
{\htwoo} rotational deexcitation within the ground vibrational state
in collisions with the {\otwo} molecule.
We filled this gap by adopting the rate coefficients for collisions with {\htwo}
and scaling them by $(\mu_{\rm{H}_2\rm{O}-\rm{H}_2}/\mu_{\rm{H}_2\rm{O}-\rm{O}_2})^{1/2}${$\approx$}0.40.
We proceeded similarly for the rotational relaxation within the vibrational state (010).
\\

The collisions with {\otwo} that involve a change in the {\htwoo} vibrational state
are of interest in the NLTE modelling of air \citep{feofilovetal2009,yankovskyetal2011,
funkeetal2012,manuilovaetal2015,langetal2020}. 
The usual scheme is similar to that of Eq. \ref{h2o_relaxation_scheme_eq}
with {X}$\equiv${\otwo}(0). 
The relaxation rate coefficient for states (100,001) has been determined by
\citet{finzietal1977}. 
The rate coefficient for {\htwoo}(010)+{\otwo}(0){$\rightarrow$}{\htwoo}(000)+{\otwo}(0)
is unclear but is expected to be much lower than that for the nearly-resonant channel 
{\htwoo}(010)+{\otwo}(0)$\leftrightarrow${\htwoo}(000)+{\otwo}(1) \citep{huestis2006}. 
Similarly, {\htwoo}(020)+{\otwo}(0)$\rightarrow${\htwoo}(010)+{\otwo}(0) is 
expected to be slower than {\htwoo}(020)+{\otwo}(0){$\leftrightarrow$}{\htwoo}(010)+{\otwo}(1).
This entails that the {\htwoo} and {\otwo} NLTE problems are coupled 
and in principle must be treated together, as is often done in the investigation of the
Earth's atmosphere.
We compiled the rate coefficients listed in Table \ref{kqvv_o2_molecule_table}, 
from which it is possible to assess the significance of collisions with {\otwo} 
in the {\htwoo} NLTE problem. 
For example, 
 Tables \ref{kqvv_h2_molecule_table}-\ref{kqvv_h2o_molecule_table}
show that the relaxation of {\htwoo}(010) in collisions with 
{\htwo} and {\htwoo} proceeds with rate coefficients on the order of 
a few$\times$10$^{-12}$
and 5.5$\times$10$^{-11}$ cm$^3$s$^{-1}$, respectively, whereas
the rate coefficient for collisions with O$_2$(0) (proceeding through
the nearly-resonant channel)
is $\sim$7$\times$10$^{-13}$ cm$^3$s$^{-1}$ at 500 K \citep[][fig. 3]{huestis2006}. 
The latter channel appears to be notably slower than the others and suggests
that unless O$_2$ is much more abundant than the other particles, 
the contribution of {\otwo} to the {\htwoo} NLTE problem should be minor. 
\\

\subsubsection*{\label{collisions_he_sec} Collisions with He.}

\citet{greenetal1993} and \citet{yangetal2013} have calculated the
rate coefficients for {\htwoo} rotational deexcitation within the ground vibrational state in collisions
with He atoms. 
We adopted the less comprehensive but more recent rate 
coefficients of \citet{yangetal2013} and 
complemented them with those of \citet{greenetal1993}. 
We extrapolated the resulting set to higher-energy states with the NM87 method.
To our knowledge, there are no calculations of rotational deexcitation rates within
the vibrational state (010), and we estimated them by scaling the rate coefficients for 
collisions with {\htwo} by 
$(\mu_{\rm{H}_2\rm{O}-\rm{H}_2}/\mu_{\rm{H}_2\rm{O}-\rm{He}})^{1/2}${$\approx$}0.74.\\

\citet{stoecklinetal2021} have calculated the  (010)$\rightarrow$(000)
rate coefficients, which compare acceptably well with the high-temperature 
measurements  of \citet{kungcenter1975}. 
We combined the thermally-averaged rate coefficients of \citet{stoecklinetal2021} (their fig. 7, bottom) 
with the branching ratios $t^{\rm{H_2}}_{ji}$ to estimate the
state-to-state (010)$\rightarrow$(000) rate coefficients. Table \ref{kqvv_he_atom_table} summarizes this.
\\

\begin{table}[h]
\caption{$k^Q_{v'v''}$: Band rate coefficients for 
deexcitation of {\htwoo} in collisions with H atoms. 
Only the o-{\htwoo} rate  coefficients are quoted; those for p-{\htwoo} are similar.
Refs.: [DA15], \citet{danieletal2015}; [CA22], \citet{cabrera-gonzalezetal2022}.
}           
\label{kqvv_h_atom_table}   
\centering                        
\begin{tabular}{c c l l l l l c}    

\hline    
$v'$ & $v''$ & \multicolumn{1}{c}{200} &  \multicolumn{1}{c}{400} & \multicolumn{1}{c}{800} & \multicolumn{1}{c}{1200} & \multicolumn{1}{c}{1500} & Source \\    
\hline    
 000 &  000 &   3.95($-$11) & 8.38($-$11) & 1.61($-$10) & 2.21($-$10) & 2.59($-$10) & [DA15]  \\
\hline
     &   &   7.10($-$11) & 1.14($-$10) & 1.89($-$10) & 2.46($-$10) & 2.82($-$10)   & $k_u^Q$($\downarrow$)  \\
\hline    
 010 &  000 &   5.93($-$14) & 1.76($-$13) & 5.24($-$13) & 9.89($-$13) & 1.56($-$12) & [CA22] \\ 
 010 &  010 &   3.96($-$11) & 8.37($-$11) & 1.61($-$10) & 2.21($-$10) & 2.57($-$10) & See text  \\

\hline
\end{tabular}
\end{table}

\begin{table}[h]
\caption{$k^Q_{v'v''}$: Band rate coefficients for {\htwoo} deexcitation in 
collisions with O atoms. Refs.: 
[ZI89], \citet{zittelmasturzo1989}; 
[FU12], \citet{funkeetal2012}.
}           
\label{kqvv_o3p_atom_table}   
\centering                        
\begin{tabular}{c c l l l l l c}    

\hline    
$v'$ & $v''$ & \multicolumn{1}{c}{200} &  \multicolumn{1}{c}{400} & \multicolumn{1}{c}{800} & \multicolumn{1}{c}{1200} & \multicolumn{1}{c}{1600} & Source \\    
\hline    

 000 &  000 &   9.38($-$11) & 1.17($-$10) & 1.57($-$10) & 1.95($-$10) & 2.24($-$10) & $\mu$ scaling \\  
 010 &  000 &   5.93($-$13) & 7.13($-$13) & 1.64($-$12) & 3.27($-$12) & 5.66($-$12) & $\mu$ scaling \\   
 010 &  010 &   8.14($-$11) & 1.30($-$10) & 1.90($-$10) & 2.26($-$10) & 2.53($-$10) & $\mu$ scaling \\
 020 &  010 &   1.05($-$12) & 1.26($-$12) & 2.92($-$12) & 5.89($-$12) & 1.03($-$11) & $\mu$ scaling \\
 100,001 &  020 &   2.45($-$12) & 3.46($-$12) & 4.90($-$12) & 6.00($-$12) & 6.93($-$12) & [ZI89,FU12] \\

\hline
\end{tabular}
\end{table}

\begin{table}[h]
\caption{$k^Q_{v'v''}$: Band rate coefficients for deexcitation of {\htwoo} in 
collisions with {\otwo}. 
References: [FI77], \citet{finzietal1977};
            [HU06], \citet{huestis2006}; 
            [FU12], \citet{funkeetal2012}. 
            $\dagger$: Rate coefficients actually refer to the nearly-resonant channels 
{\htwoo}(010)+{\otwo}(0)$\rightarrow${\htwoo}(000)+{\otwo}(1) and
{\htwoo}(020)+{\otwo}(0)$\rightarrow${\htwoo}(010)+{\otwo}(1).
}           
\label{kqvv_o2_molecule_table}   
\centering                        
\begin{tabular}{c c l l l l l c}    

\hline    
$v'$ & $v''$ & \multicolumn{1}{c}{200} &  \multicolumn{1}{c}{400} & \multicolumn{1}{c}{800} & \multicolumn{1}{c}{1200} & \multicolumn{1}{c}{1600} & Source \\    
\hline    

 000 &  000 &   8.16($-$11) &  1.02($-$10) & 1.36($-$10) & 1.69($-$10) & 1.95($-$10) & $\mu$ scaling \\
 010 &  000 &   7.00($-$13) &  7.00($-$13) & 7.00($-$13) & 7.00($-$13) & 7.00($-$13) & [HU06]$\dagger$ \\ 
 010 &  010 &   7.08($-$11) &  1.13($-$10) & 1.65($-$10) & 1.97($-$10) & 2.20($-$10) & $\mu$ scaling \\
 020 &  010 &   2.00($-$12) &  2.00($-$12) & 2.00($-$12) & 2.00($-$12) & 2.00($-$12) & [FU12]$\dagger$ \\
 100,001 &  020 &   3.30($-$13) & 3.30($-$13) & 3.30($-$13) & 3.30($-$13) & 3.30($-$13) & [FI77] \\

\hline
\end{tabular}
\end{table}

\begin{table}[h]
\caption{$k^Q_{v'v''}$: Band rate coefficients for {\htwoo}
deexcitation in collisions with He atoms. 
Only the o-{\htwoo} rate coefficients are reported; those for p-{\htwoo} are similar.
Refs.: 
[GR93], \citet{greenetal1993}; 
YA13, \citet{yangetal2013}; 
ST21, \citet{stoecklinetal2021}.
}
\label{kqvv_he_atom_table}   
\centering                        
\begin{tabular}{c c l l l c}    

\hline    
$v'$ & $v''$ & \multicolumn{1}{c}{200 K} &  \multicolumn{1}{c}{400 K} & \multicolumn{1}{c}{800 K} &  Source \\    
\hline    
 000 &  000 &   4.21($-$11) & 7.48($-$11) & 1.27($-$10) &   [GR93], [YA13]  \\
\hline
     &      &   3.51($-$11) & 7.24($-$11) & 1.37($-$10) &   $k_u^Q$($\downarrow$)  \\
\hline    
 010 &  000 &   1.81($-$15) & 1.38($-$14) & 1.00($-$13) &   [ST21] \\ 
 010 &  010 &   1.31($-$10) & 2.09($-$10) & 3.06($-$10) &   $\mu$ scaling \\  
\hline
\end{tabular}
\end{table}

\clearpage
\newpage

\end{document}